\documentclass[12pt]{article}

\usepackage{amsmath}
\usepackage{amssymb}
\usepackage{mathrsfs}
\usepackage[dvips]{graphicx}
\input epsf

\numberwithin{equation}{section}

\def\bfone{\relax{\rm 1\kern-.36em 1}}
\def\zero{\relax{ 0\kern-.38em 0}}
\def\ZZ{\relax{Z\kern-.50em Z}}

\newcommand{\CF}{ {\cal F} }
\newcommand{\CG}{ {\cal G} }
\newcommand{\CN}{ {\cal N} }
\newcommand{\CZ}{ {\cal Z} }

\newcommand{\hv}{ h^{\vee}  }
\newcommand{\av}{ \alpha^{\vee}  }
\newcommand{\AD}{ A_\mathrm{D}  }
\newcommand{\gc}{ \mathsf{g}_{\,\mathrm{C}}  }

\begin{document}

\addtolength{\baselineskip}{2pt}
\thispagestyle{empty}

\vspace{2.5cm}

\begin{center}
{\scshape\Large 
Supersymmetric $\CN=2$ gauge theory with arbitrary gauge group}

\vspace{1.5cm}

{\scshape\large Michael Yu. Kuchiev}

\vspace{0.5cm}
{\sl School of Physics, University of New South Wales,\\
Sydney, Australia}\\
{\tt kmy@phys.unsw.edu.au}\\\vspace{1cm}

{\Large ABSTRACT}

\vspace{0.3cm}

\end{center}

\begin{abstract}
A new universal model to implement the Seiberg-Witten approach to low-energy properties of the supersymmetric $\CN=2$ gauge theory with an arbitrary compact simple gauge group, classical or exceptional, is suggested. It is based on the hyperelliptic curve, whose genus equals the rank of the gauge group.  The weak and strong coupling limits are reproduced. The magnetic and electric charges of light dyons, which are present in the proposed model comply with recent predictions derived from the general properties of the theory. The discrete chiral symmetry is implemented, the duality condition is reproduced, and connections between monodromies at weak and strong coupling are established. It is found that the spectra of monopoles and dyons are greatly simplified when vectors representing the scalar and dual fields in the Cartan algebra are aligned along the Weyl vector. This general feature of the theory is used for an additional verification of the model. The model predicts the identical analytic structures of the coupling constants for the theories based on the SU($r+1$) and Sp($2r$) gauge groups.
\end{abstract}

`
%\begin{keyword}
\\
Keywords: Gauge theory, supersymmetry, N=2, Seiberg Witten solution\\
PACS 11.15.-q, 11.30.Pb
%\end{keyword}
%\end{frontmatter}

% main text
\newpage

\section{Introduction}
\label{intro}

The low-energy solution for the $\CN=2$ supersymmetric gauge theory
found by Seiberg and Witten \cite{Seiberg:1994rs,Seiberg:1994aj} for the SU(2) gauge group was generalized to cover other gauge groups in 
%[3-23].  
\cite{Klemm:1994qs,%           SU(n)
Argyres:1994xh,%         SU(n)    
Klemm:1994qj,%           SU(n)
Douglas:1995nw,%         SU(n)
Danielsson:1995is,%      SO(2n+1)
%Gorsky:1995zq,%
%%%%% Argyres:1995jj,%         new phenomena SU(3)
Hanany:1995na,%          SU(n)+quarks in fundamental
Argyres:1995wt,%         SU(n))+quarks in fundamental
Klemm:1995wp,%           SU(3)
Brandhuber:1995zp,%      SO(2n)
%Lerche:1995xj,%
Minahan:1995er,%         SU(3)+matter
Gauntlett:1995fu,%       SU(2)+matter
Argyres:1995fw,%         SO(n), Sp(2n) + matter
Hanany:1995fu,%          SO(n) + matter
%Sonnenschein:1995hv,%
%Argyres:1995xn,%
%Danielsson:1995zi,%
Alishahiha:1995wm,%      G2
%Itoyama:1995uj,%
%Itoyama:1995ca,%
%Ito:1996qj,%
%Ito:1996sq,%
%Ohta:1996fr,%
Krichever:1996ut,%      SU(n) + matter
Abolhasani:1996ik,%     Lie groups
%Dorey:1996bf,%
D'Hoker:1996nv,%        SU(n)+matter
Landsteiner:1996ut,%    G2 nonelliptic
D'Hoker:1996mu,%        SO(n), Sp(2n)+matter
%D'Hoker:1996ph,%        prepotential
%D'Hoker:1999ft,%        lectures
%Gorsky:2000ej,%
%Vainshtein:2000cka,%
2002tpet.conf....1D,%   lectures
Alishahiha:2003hj%      G2
%Janik:2005sk,%
%Janik:2007es%
}.
For the classical groups ($A,B,C,D$ series)
% $A_{n},B_{n},C_{n}$ and $D_{n}$ 
the algebraic curve, which describes the solution is believed to be hyperelliptic, though Ref.\cite{Martinec:1995by} suggested a non-hyperelliptic description for all gauge groups, which is based on the analogy with the integrable systems. Exceptional groups ($G_2,F_4$ and three $E$ groups) 
% $G_2,F_4,E_6,E_7$ and $E_8$ 
proved more challenging for an analysis, see discussion in \cite{Abolhasani:1996ik,Landsteiner:1996ut,Alishahiha:2003hj}. 

Different aspects of the problem for unitary groups SU($r$+1) were revealed in
Refs. [3-6,8-10,12,13,17-19].
Refs. 
\cite{
Klemm:1994qs,%  A         SU(n)
Argyres:1994xh,%         SU(n)    
Klemm:1994qj,%           SU(n)
Douglas:1995nw,%         SU(n)
Hanany:1995na,%          SU(n)+quarks in fundamental
Argyres:1995wt,%         SU(n))+quarks in fundamental
Klemm:1995wp,%           SU(3)
Minahan:1995er,%         SU(3)+matter
Gauntlett:1995fu,%       SU(2)+matter
Krichever:1996ut,%       SU(n) + matter
Abolhasani:1996ik,%	     lie groups
D'Hoker:1996nv%         SU(n)+matter
}.
For orthogonal SO(2$r$+1) and SO(2$r$), as well as for simplectic Sp(2$r$) gauge groups 
the problem was discussed in
Refs. 
\cite{
Danielsson:1995is,%      SO(2n+1)
Brandhuber:1995zp,%      SO(2n)
Argyres:1995fw,%         SO(n), Sp(2n) + matter
Hanany:1995fu,%          SO(n) + matter
Abolhasani:1996ik,%       lie groups
D'Hoker:1996mu%        SO(n), Sp(2n)+matter
}.
A brief summary is given in Refs. \cite{Abolhasani:1996ik,2002tpet.conf....1D%lectures
}, additional references can be found in \cite{Kuchiev:2008mv}.
Along with impressive progress, these works show issues, which need more attention. One of them is related to the fact that even for classical series of the gauge groups 
the Seiberg-Witten approach has been implemented in different forms for different gauge groups. For exceptional gauge groups the situation looks even more complicated; there seem to be no consensus as to whether the curve, which describes the solution, is hyperelliptic or not, and how it may, or may not differ qualitatively from the case of classical gauge groups. 
This situation puts restrictions on the understanding of general, basic properties of the $\CN=2$ supersymmetric theories.

Addressing this issue, the present work suggests a universal description of the low energy properties of the $\CN=2$ supersymmetric gauge theory for an arbitrary compact simple gauge group. The charges of light dyons, which were derived in \cite{Kuchiev:2008mv} 
from the basic properties of the theory, play an essential role in the presented analysis. 
Previously the strategy used for implementing the Seiberg-Witten approach was different. One had first to derive a solution and only then one could try to establish the values of the charges. However, the entire procedure was quite sophisticated and prior to \cite{Kuchiev:2008mv} the charges had been explicitly known only for several, most simple gauge groups. In the present work the transparent form of the charges 
found in \cite{Kuchiev:2008mv} is used to reveal that the low energy solution has a simple and general form.

%
%Sections \ref{SW-R} - \ref{minimal-set} give a summary of basic facts related to the Seiberg-Witten solution \cite{Seiberg:1994rs,Seiberg:1994aj}, the Riemann bilinear relations, and dyon charges \cite{Kuchiev:2008mv}, while Sections \ref{differential}-\ref{Summary} present main results of the present work.

%%

\section{$\CN=2$ supersymmetric gauge theory}
\label{SW-R}

Recall the most important for us properties of the supersymmetric  $\CN=2$ gauge theory. The theory describes the scalar field $A$, two chiral spinors $\psi$ and $\lambda$, and the gauge field $v_\mu$,  all in the adjoint representation of a gauge group, which is a simple Lie group $\mathsf{G}$ \cite{Sohnius:1985qm}.  The energy of the scalar field turns zero provided this field has a coordinate independent vacuum expectation value that lies in the Cartan subalgebra $\gc$ of the gauge algebra $\mathsf{g}$, $A\in \gc \subset \mathsf{g}$, and whose real and imaginary parts satisfy $\mathrm{Re} \,(A)\propto \mathrm{Im}\,(A)$.  Such a scalar field can be treated as an $r$-dimensional vector, which is characterized by its expansion coefficients $A\equiv A_i$, $i=1,\dots r$ in some basis. Presence of this vacuum expectation value makes the vacuum state degenerate with the moduli space given by $\gc$. The gauge symmetry is spontaneously broken, generically down to $r$ products of the gauge $U(1)$, $G \rightarrow U(1)\times \cdots \times U(1)$, where $r$ is the rank of the algebra $\mathsf{g}$. There remains also unbroken a discrete group of gauge transformations, which comprises the Weyl group of $\mathsf{g}$. In the perturbation theory regime this breaking generates masses for all degrees of freedom, except those that correspond to $r$ unbroken $U(1)$ gauge symmetries, which describe $r$ massless gauge bosons and their superpartners.

The low-energy properties of the theory are described by the prepotential $\CF$, which is a holomorphic function of the scalar field $\CF=\CF (A)$, as was argued in \cite{Seiberg:1988ur}. First derivatives of the prepotential define an $r$-vector of the dual scalar field $A_\text{\it D}$
\begin{equation}
	A_{\text{\it D},\,i}\,=\,\frac{\partial \CF}{\partial A_i}~,\quad  \quad i=1,\dots r~.
	\label{AD}
\end{equation}
Its second derivatives give the $r\times r$ matrix $\tau$ of effective coupling constants 
\begin{equation}
	\tau_{ij}\,=\,\frac{\partial A_{\text{\it D},\,i}}{\partial A_j}\,=\,\frac{\partial^2\CF}{\partial A_i\,\partial A_j}~,\quad\quad i,j=1,\dots r~.
	\label{tauF2}
\end{equation}
Real and imaginary parts of this matrix are related to the $r\times r$ matrix ${\mathrm g}$ of proper coupling constants  \footnote{Notation used for the coupling constants $\mathrm g$ should not be confused with the genus of the Riemann surface $g$ and the gauge algebra $\mathsf{g}$.} and matrix of theta-angles $\theta$,  
\begin{equation}
	\tau\,=\,\frac{\theta}{2\pi} + 4\pi i\,{\mathrm g}^{-2}~.
	\label{gij}
\end{equation}
The duality, which is key to the Seiberg-Witten approach, presumes that the couplings of dyons, which appear in the theory, are described by the dual coupling matrix
%\begin{equation}
$	\tau_\text{\it D}\,=\,-\tau^{-1}$.
Equation (\ref{tauF2}) implies that $\tau$ is a symmetrical matrix, 
while Eq.(\ref{gij}) shows that its imaginary part is a positive definite matrix
\begin{align}
	 \tau\,=\,\tau^{\,T}~,\quad\quad 	\mathrm{Im}~\tau\,>\,0~.
	\label{t=T} 
\end{align}
The latter property is presented here in a symbolic form,  
which means that all eigenvalues of the matrix in question are positive. 

For a strong scalar field the coupling is weak. Correspondingly, the quantum corrections for the scalar field are weak as well, and the field $A$ is close to its classical value $a$,
\begin{equation}
	A\simeq a~.
	\label{Aweak}
\end{equation}
The dual field in this region reads
\begin{equation}
A_\mathrm{D} \, \approx \, \frac{i}{2\pi} \, \hv\,A \, \ln \frac{A^2}{\Lambda^2}~.	
	\label{APTh2}
\end{equation}
where $\hv$ is the dual Coxeter number of the algebra.
For relevant properties of simple Lie algebras see e.g. \cite{Bourbaki:2002,Di-Francesco:1997}.
%\cite{Bourbaki:2002,Di-Francesco:1997,Slansky:1981yr,Georgi:1982jb}. 
\footnote{The dual Coxeter number  $\hv=\hv(\mathsf{g})$ of the algebra $\mathsf{g}$, the eigenvalue of the quadratic Casimir operator in the adjoint representation $C_2(\mathsf{g})$,
and the Dynkin index of the adjoint representation $\chi_\mathrm{adj}(\mathsf{g})$ are all related, $2\hv(\mathsf{g})=C_2(\mathsf{g})=2\chi_\mathrm{adj}(\mathsf{g})$,  see e.g. \cite{Di-Francesco:1997}, Eqs.(13.128),(13.134).} 
Eq.(\ref{APTh2}) is written in the large-logarithm approximation, in which the argument of the logarithmic function is approximated as $\propto A^2$ thus neglecting its possible dependence on the direction of the vector $A$.
Eq.(\ref{APTh2}) implies that the Gell-Mann - Low beta-function for weak coupling  is characterized by the coefficient 
$	b^{(\text{w})}\,=\,2 \hv$, 
as it should for the $\CN=2$ supersymmetric gauge theory, see e.g. Ref.\cite{Peskin:1997qi}.

The theory possesses the chiral symmetry. On the classical level it manifests itself via the  continuous transformation of the fields
\begin{equation}
%\label{U(1)}  
 \vartheta \rightarrow e^{i\gamma} \vartheta~,\quad
	\psi  \rightarrow e^{i\gamma} \psi~, \quad
	\lambda \rightarrow e^{i\gamma} \lambda~,\quad  
	A \rightarrow e^{2i\gamma} A~.
	\label{2gammaA}
\end{equation}
Here $\vartheta$ is a conventional anti-commuting variable of the $\CN=1$ superspace \cite{Wess:1991}. Quantum corrections break this symmetry to $Z_{\,4\hv}$; the phase $\gamma$ takes only discrete values
\begin{equation}
	\gamma=2\pi\,\frac {m} {4\hv}~,\quad m=0,1,\,\dots \,4 \hv-1~.
	\label{gamma}
\end{equation}
The effect, which leads to this restriction on $\gamma$ is related to the variation of the $\theta$-angle of the theory, which takes place due to the chiral transformation and  reads 
%\begin{equation}
$\Delta \,\theta \,=\,4\hv\gamma$. 
The chiral symmetry persists provided that the variation of $\theta $  is an integer of $2\pi$, $ \Delta \,\theta\,=4\hv\gamma=\,2\pi m$, which justifies Eq.(\ref{gamma}).

The transformation of the scalar field $A$ in Eq.(\ref{2gammaA}) is accompanied by the transformation of the dual field $\AD$. Using Eq.(\ref{APTh2}) one finds 
\begin{equation}
\AD\rightarrow \AD^{\,\prime}\,= \,
%e^{2i\gamma} \left(\AD-\frac{\gamma}{\pi}\,\sum_\alpha\,(\alpha\cdot A)\alpha \right)	=
\exp\left( \pi i/\hv \right) \left(\AD-A \right).	
	\label{ADZ}
\end{equation}
Here the second term in the brackets originates from the logarithmic function in 
Eq.(\ref{APTh2}), and  it is assumed that $m=1$ in Eq. (\ref{gamma}).
Eq.(\ref{ADZ}) shows that the defining element of the chiral ${Z}_{\,4\hv}$ symmetry  manifests itself via the following transformation of $\Phi$
\begin{equation}
\Phi\,\rightarrow \,
\Phi^{\,\prime} \, = \,
\exp\left( \,\pi i/\hv \right) H\,\Phi~,
	\label{ADgamma}
\end{equation}
where $\Phi$ is the $2r$ vector of the fields
\begin{equation}
\Phi\,=\,
\begin{pmatrix} A_\text{\it D} \\ A\end{pmatrix}~,
\label{Phi}
\end{equation}
and $2r\times 2r$ times matrix $H$ can be written as the following $2\times 2$ block matrix
\begin{equation}
H\,=\,\begin{pmatrix}~ 1 & \!-1 \\ ~0 & \,~\,1 \end{pmatrix}~.	
\label{Mch}
\end{equation}
Eq.(\ref{ADgamma}) can be considered a monodromy that arises when the phase $\gamma$  is treated as a continuous variable that varies from $\gamma=0$ to the value $\gamma=2\pi/\hv$ allowed by Eq.(\ref{gamma}). 

%\section{Duality}
%\label{Duality}
It was explained in \cite{Seiberg:1994rs} that the theory satisfies the important condition of duality, which amounts to the following transformation of the fields
\begin{align}
&\Phi\rightarrow \Phi^\prime\,=\,\Omega\,\Phi~,
\label{dual}
\\
&\Omega\,=\,\begin{pmatrix} ~~0 & 1~\\
													   -1 & 0~
													\end{pmatrix}~,
													\label{Omega}
\end{align}
that keeps the description of the theory intact. 
%In the detailed notation it  reads $A\rightarrow-A_\text{\it D}$, $A_\text{\it D}\rightarrow A$, and $\tau\rightarrow\tau_\text{\it D}=-\,\tau^{-1}$.

The dyons in the Seiberg-Witten approach are BPS states \cite{Bogomolny:1975de,Prasad:1975kr}. Consequently, 
as was noted in \cite{Witten:1978mh}, the mass $m_{\CG}$ of a dyon is related to the central charge $\CZ_{\,\CG}$
\begin{align}
\label{MZ}
	 m_{\,\CG}   & \,=\,  2^{\,1/2} \,\,|\,\CZ_{\,\CG}\,|~, \\ 
\label{Z}	 
	 \CZ_{\,\CG}     &   \,=\,   g \cdot \AD+q\cdot A \,\equiv \,\CG \,\varPhi~.
	\end{align}
Here the dyon charge $\CG$ is defined as follows
\begin{equation}
 \CG\,=\,(\,g,\,q\,)~.
%& \Phi\,=\,
%\begin{pmatrix} A_\text{\it D} \\ A\end{pmatrix}~.
%\label{Phi}
\end{equation}
It was argued in \cite{Kuchiev:2008mv} that properties of light dyons can be described using a set of dyons, whose magnetic and electric charges $g,q$ satisfy 
\begin{equation}
	\CG_{\,\alpha_i, \,m}\,\equiv\,(g,q)\,=\,(\,\av_i,-m\,\av_i\,)~.
	\label{ma}
\end{equation}
Here $\av_i$, $i=1,\dots r$, is a set of simple coroots, and $m\,\in\,Z_{\,\hv}$.  
Eq.(\ref{ma}) can be supported by simple physical arguments. The charge of the monopole should be a vector that belongs to the dual lattice, as follows from the Dirac-Schwinger-Zwanziger quantization condition. The simplest possible charge of the monopole should therefore be 
\begin{equation}
(g,q)=\CG_{\,\alpha_i, \,0}=(\av_i,0)~, 
\label{monopole}
\end{equation}
where $\av_i$ is a simple coroot. Using the Witten effect, which states that a presence of the magnetic charge $g$ in the $\theta\ne 0$ vacuum ($\theta$ is the conventional theta-angle) leads to the presence of the electric charge, one verifies then that the chiral transformations (\ref{ADgamma}) convert the monopole with the charge $\CG_{\,\alpha_i, \,0}$ into dyons, whose charges are $\CG_{\,\alpha_i, \,m}$. 
It was stated in \cite{Seiberg:1994rs} that there is a possibility for the condensation of monopoles or dyons, which results in explicit breaking of the $\CN=2$ supersymmetric gauge theory down to $\CN=1$ gauge theory. In order to describe this transition it suffices to have only $\hv$ dyons for a given $\av_i$. This implies that the integer $m$ in Eq.(\ref{ma}) should be taken modulo $\hv$.

\section{Periods and fields}
\label{Periods and fields}
We will see that a very convenient way to describe the scalar field $A$, its dual $A_\text{\it D}$ and the classical value of the scalar field $a$  provides the basis of the fundamental weights 
\begin{align}
&A ~\, = \,\sum_{i=1}^{r}~A_i~\omega_i~.
\label{Aq}
\\
&A_\text{\it D} = \,\sum_{i=1}^{r}\,A_{\text{\it D},\,i}~\omega_i~,
\label{ADq}
\\
&~a ~\, = \,\sum_{i=1}^{r}~a_i~\omega_i~.
\label{ome}
\end{align}
Here $\omega_i$ are the fundamental weights of the gauge algebra, while the coefficients ${ A}_i$, ${ A}_{\text{\it D},\,i}$ and  $a_i$ represent the fields in the chosen basis.
The expansion coefficients of the fields introduced in Eqs.(\ref{Aq})-(\ref{ome}) can be expressed via the fields using the scalar product in the gauge algebra. Taking for example the field $A$ one can state that 
%\begin{equation}
$A_j\,=\,A\cdot\av_j$,
%\label{aav}
%\end{equation}
where the fact that the simple coroots $\av_i$ and fundamental weights $\omega_j$ are mutually orthonormal, 
%see e.g.  \cite{Bourbaki:2002},
%%Di-Francesco:1997,Slansky:1981yr,Georgi:1982jb},
\begin{equation}
\av_i\cdot\omega_j\,=\,\delta_{ij}~,
\label{avo}
\end{equation}
was used. 
%\footnote{In the present work  the subscripts enumerate the vectors,  $\alpha_i$ and $\omega_j$  refer to the $i$-th simple root and $j$-th fundamental weight, $i,j=1,\dots r$, while the superscripts indicate their components in orthogonal basis. These arrangements differ from Ref. \cite{Kuchiev:2008mv} where subscripts mark the components of vectors.}
The basis of fundamental weights, which is used for the fields in (\ref{Aq}), (\ref{ADq}) greatly simplifies expressions for the central charges $\CZ_{\,\CG}$ (\ref{Z}). For example, for the dyons with the electric and magnetic charges $	\CG_{\,\alpha_i, \,m}$ from (\ref{ma}) one finds using (\ref{avo})
\begin{equation}
	 \CZ_{\,\CG_{\,\alpha_i, \,m} }    \,=\,A_{\,i}-m\,A_{\text{\it D},\,i}~.
\label{A-mAD}
\end{equation}
Further confirmation of convenience of the basis of the fundamental  weights is unfolded later, when the strong coupling limit in Section \ref{Strong-coupling} is discussed. 
%Here let us only briefly mention that this basis provides the easiest way to ensure that the dyons acquire the necessary charges (\ref{ma}).  
Eqs.(\ref{Aq}), (\ref{ADq}) imply that the prepotential is considered a function of the coefficients $A_i$,
\begin{equation}
\CF\,=\,\CF(A_1,\dots,A_r)~, 
\label{FFA}
\end{equation} 
which specifies the precise meaning of Eqs. (\ref{AD}), (\ref{tauF2}).

Following the spirit of the Seiberg-Witten approach  define the expansion coefficients in Eqs.(\ref{Aq}), (\ref{ADq}) using properties of a Riemann surface. Presume that on the Riemann surface there is the differential $d\lambda$, which is a holomorphic function of $r$ parameters. Presume also that on this surface there exists a set of cycles $C_i$ and $C_{\text{\it D},\, i}$, $i=1,\dots r$, which interception form $( \,C\,|\,C^{\,\prime}\, )$ is canonical
\begin{align}
	&(C_i|\,C_j)\,=\,(C_{\text{\it D},i}\,|\,C_{\text{\it D},j})\,=\,0~,
	\label{CC} \\
	&( C_i\,|\,C_{\text{\it D},j})\,=\,-( C_{\text{\it D},j}\,|\,C_i )\,=\,\delta_{ij}~.
	\label{CC=d}
\end{align}	
The periods that correspond to these cycles are identified with the scalar and dual fields 
\begin{align}
	& { A}_i ~~~=~\frac{1}{2\pi i}~\oint_{C_i}~d\lambda~,
	\label{Ai2} \\
	& { A}_{\text{\it D},\,i}\,=\,\frac{1}{2\pi i}~\oint_{C_{\text{\it D},\,i}}\!\!\!\!d\lambda~.
	\label{ADi2}
\end{align}
This identification ensures that the basic properties of the $\tau$ matrix described by  (\ref{t=T}) are satisfied automatically due to the Riemann bilinear relations. The necessary Riemann surface, the differential $d\lambda$ and the set of cycles $C_{i}$ and $C_{\text{\it D},\,i}$ are all defined below in Section \ref{differential}.

\section{Differential and Riemann surface}
\label{differential}

Consider the following differential
\begin{equation}
	d\lambda\,=\,\sqrt{z}~\frac{X(z)}{Y(z)}~dz~.
	\label{dx}
\end{equation} 
Here the complex variable $z$ is defined on the Riemann surface specified below,
$X(z)$ and $Y(z)$ are holomorphic functions of $z$. Take the function $Y(z)$ as the following hyperelliptic curve 
\begin{equation}
	Y^2(z)\,=\,P^2(z)-Q^2~,
	\label{Y}
\end{equation}
where $P(z)$ is a polynomial of $z$ and $Q$ is a constant. 
Discussion of the weak coupling in Section \ref{weakcoupling} will justify that the nominator $X(z)$ in Eq.(\ref{dx}) should be written as follows
\begin{equation}
	X(z)\,=\,\frac{dP(z)}{dz}~.
	\label{X}
\end{equation}
Presume that the genus of the Riemann surface, which is associated with the differential (\ref{dx}) should be equal to the rank $r$ of the gauge group. Then the polynomial $P(z)$ 
should have the power $r$, having thus the form
\begin{equation}
P(z)\,=\,\prod_{i=1}^{r}\,(z-a_i^2)~,
\label{P(z)}
\end{equation}
where $a_i^2$,  $i=1,\dots r$ are its nodes.
Discussing the perturbation theory in Section \ref{weakcoupling} we will see that these nodes equal the coefficients of the expansion of the classical scalar field $a$ over the basis of fundamental weights in (\ref{ome}).
%The convenience of the basis of the fundamental weights for $A$ becomes clear in another limiting case, namely for the strong coupling, see Section \ref{Strong-coupling}. 
%Setting out to present the field $A$ in the basis of the fundamental weights, we have to consider its classical analogue $a$ in (\ref{ome}) in the same basis.

The definition (\ref{P(z)}) for  $P(z)$ is closely related to the classical description of the scalar field. The quantum effects are brought into the curve Eq.(\ref{Y}) via $Q$, which should necessarily depend on $\Lambda$. We presume that 
\begin{equation}
Q\,=\,[ \,a\, ]^{\,2r-\hv}\,\Lambda^{\hv}~.
\label{Q(z)}
\end{equation}
Here the symbol $[ \,a\, ]$ is defined as follows
\begin{equation}
[ \,a\, ]^{\,2r}\,=\,\frac{1}{r}~\sum_{i=1}^r\,(a_i)^{\,2r}~.
\label{f2}
\end{equation}
The condition $Q\propto \Lambda^{\hv}$ in Eq.(\ref{Q(z)}) becomes clear from discussion of the perturbation theory in Section \ref{weakcoupling}, which shows that this power of $\Lambda$ reproduces the necessary coefficient $b^{(\text{w})}=2\hv$ of the Gell-Mann - Low beta-function. The power $2r-\hv$ of the field $a$ in Eq.(\ref{Q(z)}) for $Q$ follows from simple dimensional counting. The precise form (\ref{f2}) in which the field $a$ appears in $Q$ is inspired by two reasons. First, $[ \,a\, ]$ needs to be an even function of all $a_j$, $j=1,\dots r$, which makes the curve $Y(z)$ an even function of these parameters as well. %Advantages of this property of the curve are discussed in Section \ref{Weyl+symmetry}. 
To see the second reason, consider a particular value of the classical field $a$, which guarantees that
$Y^2(z)$ has the $r$-times degenerate node at $z=0$,  i.e.  $Y^2(z)=O(z^r)$ when $z\rightarrow  0$. Each node of $Y^2(z)$ located at $z=0$ makes one monopole massless, see Section \ref{Strong-coupling}. Consequently, the $r$-degenerate node makes $r$ different monopoles massless. When $r$ massless monopoles are present one can consider an explicit breaking of the supersymmetry from $\CN=2$ down to $\CN=1$ case, as suggested by Seiberg and Witten  \cite{Seiberg:1994rs}. 
This phenomenon underlines an important role played in the theory by the degenerate node of $Y^2(z)$ at $z=0$. Equation (\ref{f2}) makes 
it certain that this degenerate node is present, thus  paving the way for the phenomenon of breaking of the $\CN=2$ supersymmetry down to $\CN=1$.
%\footnote{
%It can be verified that the definition of $[ \,a\, ]$ in Eq.(\ref{f2}) allows one to choose a classical field $a$ in such a way as to reduce the curve to $Y^2(z)=z^r(z^r+const)$. An attempt at a different definition of $[ \,a\, ]$, for example the simplest that comes to mind  $[ \,a\, ]^2\propto \sum_{i=1}^r(a_i)^2$, or more sophisticated $[ \,a\, ]^{\,2k}\propto \sum_{i=1}^r(a_i)^{2k}$ with $k<r$, would not allow $Y^2(z)$ to have the necessary multiple pole at $z=0$.} 
This issue will be developed in more detail elsewhere. The factor $1/r$ is added in the right-hand side of Eq.(\ref{f2}) to simplify calculations.

Equations (\ref{P(z)}),(\ref{Q(z)}) define the curve $Y(z)$ in Eq.(\ref{Y}), which is turn defines the differential $d\lambda$ in Eq.(\ref{dx}). Summing up, we write the differential
\begin{align}
&d\lambda\,=\,\sqrt{z}~\,\frac{P^{\,\prime}(z)}{Y(z)}~\,dz~,
\nonumber
\\
&Y^2(z)\,=\,P^2(z)-Q^2~,
\label{YY}
\\
&P(z)\,=\,\prod_{i=1}^{r}\,(\,z-a_i^2\,)~,\quad \quad Q\,=\,
[ \,a\,]^{\,2r-\hv}\,\Lambda^{\hv}~.
\nonumber
\end{align}
For some applications below it is natural to work with slightly modified expressions for the differential. One such modification, which is convenient for the weak coupling limit discussed in Section \ref{weakcoupling}, reads
\begin{align}
&d\lambda\,=\,\sqrt{z}~\frac{dR ~}{ \,\left( R^2-1 \right) ^{1/2} }~,
	\label{dR}
\\
&R(z)\,=\,{P(z)}/{Q}~,
	\label{R(x)}
\end{align}
Another useful set of variables is based on the transformations $z\rightarrow \zeta$, $a_i\rightarrow \xi_i$, 
$Y(z)\rightarrow y(\zeta)$, 
$P(z)\rightarrow p(\zeta)$,
$Q\rightarrow q$, which allow the differential (\ref{YY}) to be rewritten as follows
\begin{align}
&\zeta \,=\, z/[ \,a\, ]^2~,\quad\quad \xi_i\,=\,a_i/[ \,a\, ]~,
\label{zeta}
\\
&d\lambda\,=\,[ \,a\, ]\,\sqrt{\zeta}~~\frac{ p^{\,\prime}(\zeta)}{y\,(\zeta) }~\,d\zeta~,
\label{scale}
\\
&y^2\,(\zeta)\,=~p^2\,(\zeta)- q^2~,
%=\,Y^2(z)/a^{4r}\,
\label{yzeta}
\\
&p\,(\zeta)\,=\,\prod_{i=1}^{r}\,(\zeta-\xi_i^2)~,\quad~q\,=\,\Lambda^{\hv}\! /\, [ \,a\, ]^{\,\hv}~.
%\,=\,P(z)/a^{2r}
\label{p}
\end{align}
Equations Eqs.(\ref{YY})-(\ref{p}) propose that the functional form of the differential $d\lambda$ is universal, is equally applicable for all gauge groups, classical, exceptional, simply-laced or not. There are two governing parameters the rank $r$ and  Coxeter number $\hv$ of the gauge group.
For convenience they are listed in Table \ref{t1} for all compact simple Lie groups. 
Observe that the simply-laced SU($r+1$) and non-simply laced Sp($2r$) have same rank and same Coxeter number. Important implications of this fact are discussed in Section \ref{GG}.
\begin{table}[h]
%\centering
		\begin{tabular}{|l|c|c|c|c|c|c|c|c|c|}
		
		\hline
		
Group & SU($r+1$) & SO($2r+1$) & Sp($2r$) & SO($2r$) & G$_2$ & F$_4$ & E$_6$ 
& E$_7$  & E$_8$ \\
     
			\hline
	
Class    & $A_r$    & $B_r$      & $C_r$    &  $D_r$   &       &       &      &      & \\

			\hline
					
~~~ $r$ & $r$     & $r$ & $r$  &  $r$     &    2  & 4     &  6     & 7     & 8      \\

			\hline

~~~$\hv$  & $r+1$   & $2r-1$& $r+1$  &  $2r-2$     &  4  & 9     &  12     & 18     & 30   \\

			\hline			
\end{tabular}
\vspace{0.3cm}
\caption{The rank $r$ and dual Coxeter number $\hv$ of compact simple Lie groups.}
		\label{t1}
		\end{table}
%\section{Riemann surface}
%\label{RS}
%\noindent

\begin{figure}[tbh]
  \centering \includegraphics[ height=5cm, keepaspectratio=true, angle=0]{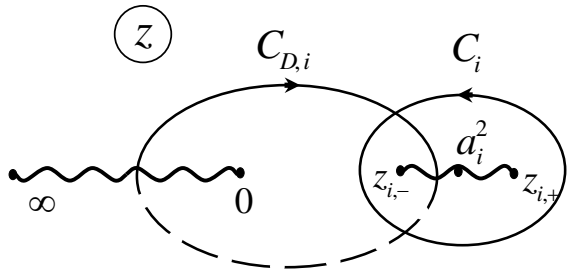}
\vspace{0cm}
\caption{ The Riemann surface related to the differential $d\lambda$ in Eq.(\ref{YY}).
The wavy lines show the cuts. One runs between the nodes $z_{i,\,\pm}$ of the curve $Y(z)$; 
there exist $r$ pairs of nodes with the cuts between them, only one is shown here. A cut due to the factor $\sqrt{z}$ in $d\lambda$ runs between $z=0$ and $\infty$. The cycle $C_i$ lies on the first sheet, the cycle $C_{\text{\it D},\,i}$ runs over the first and second sheets of the Riemann surface where it is shown by solid and dashed lines respectively. }\label{FIG1} 
   \end{figure}

The singularities of the differential in Eq.(\ref{YY}) originate from two sources. One is the factor $\sqrt{z}$. In order to classify others, which come from the nodes of the curve $Y^2(z)$,  it is convenient to present the curve as follows
\begin{align}
&Y^2(z) \,=\,Y_+(z)\,Y_-(\zeta)~,
\label{YYY}
\\
&Y_\pm(z)\,=\,P(z)\,\pm \,[\,a \,]^{\,2r-\hv}\Lambda^{\hv}~.
\label{Ypm}
\end{align} 
Since the order of the polynomial $P(z)$ is $r$, there are $r$ nodes of $Y_+(z)$,  and $r$ nodes of $Y_-(z)$. Call the former $z_{i,+}$ and the latter $z_{i,-}$, so that 
\begin{equation}
Y_{\,\pm}\,(z_{\,i,\pm})\,=\,0~, \quad\quad i=1,\dots r~.
\label{zpm}
\end{equation}
One can presume that all the nodes are enumerated here in such a way as to make it certain that the pairs of nodes $z_{\,i,\,+}$ and $z_{\,i,\,-}$ coincide in the limit  $\Lambda=0$. 
To construct the Riemann surface for the differential in Eq.(\ref{YY}) we put $r$ cuts on the complex plane $z$, each cut connecting a pair of the nodes $z_{i,\,+}$ and $z_{i,\,-}$. One more cut, which runs from $0$ to $\infty$, is prompted by the function $\sqrt{z}$. Clearly, the Riemann surface, which arises after the appropriate gluing is taken, represents the hyperelliptic structure illustrated by Fig. \ref{FIG1}, where for simplicity only one pair of nodes $z_{i,\,\pm}$ is shown. 
   \noindent
The genus of this Riemann surface $g$ is defined by the number of cuts. The simple counting, see e. g. page 212 of \cite{Springer:1957}, shows that in our case the equality holds between the genus of the Riemann surface and rank of the gauge group 
\begin{equation}
g\,=\,r~.
\label{g=r}
\end{equation}

\section{SU(2) gauge theory}
As an illustration consider the gauge theory with the SU(2) gauge group, which was discussed in \cite{Seiberg:1994rs}. Taking the values  $r=1$, $\hv=2$, $\omega_1=1/\sqrt{2}$, which are valid for the SU(2) gauge group,
one finds $P(z)=(z-a^2_1)$, $a_1=\sqrt{2}a$, $Q=\Lambda^2$, $Y^2(z)=(z-a^2_1)-\Lambda^4$. Substituting this into  Eqs.(\ref{Ai2})-(\ref{ADq}) one writes
\begin{align}
&A_1~~\,=~\frac{1}{2\pi i}~\oint_{C_1} ~\frac{ \sqrt{z}~\,dz }{ (\,(\,z-a^2_1\,)^2-\Lambda^4\,)^{1/2} }~,
\\
&A_{\text{\it D},\,1}\,=\,\frac{1}{2\pi i}~\oint_{C_{\text{\it D},\,1}}\, \frac{ \sqrt{z}~\,dz }{(\,(\,z-a^2_1\,)^2-\Lambda^4\,)^{1/2} }~,
\end{align}
where $C_1$ and $C_{\text{\it D},\,1}$ are defined in accord with Fig. \ref{FIG1}, and the coefficient $1/\sqrt 2$ in front of the integrals arises from the factor $\omega_1$ in Eqs.(\ref{Aq}),(\ref{ADq}). Using a new integration variable $x=(a^2_1-z)/\Lambda^2$ and introducing a parameter $u=a^2_1/\Lambda^2=2a^2/\Lambda^2$, one rewrites the integrals in the form
\begin{align}
&A~~\,\,=~\frac{\Lambda}{\pi\sqrt 2}\,\int_{-1}^1\, \frac{ \sqrt{x-u} }
{\sqrt{x^2-1 }}~dx\,,
\label{A1}
\\
&A_{\text{\it D}}\,=~\frac{\Lambda}{\pi\sqrt 2}\,\int_{\,1}^{\,u}
\frac{ \sqrt{x-u} }
{\sqrt{x^2-1 }}~dx~,
\label{AD1}
\end{align}
which agrees with \cite{Seiberg:1994rs}.

\section{Solution at weak coupling}
\label{weakcoupling}
\subsection{Scalar field inside the Weyl chamber}
\label{inside Weyl chamber}

Consider the weak coupling limit, when the scalar field is large 
\begin{equation}
\Lambda^2/a^2\rightarrow 0~. 	
	\label{L0}
\end{equation}
Then using Eqs.(\ref{YY}) one finds that the nodes of the curve $z_{i,\,\pm}$ introduced in Eq.(\ref{zpm}) satisfy
\begin{equation}
z_{i,\,+}\,\approx \, z_{i,\,-} \approx a_i^2~,
\label{zpmapp}
\end{equation}
while the function $R=R(z)$ in Eq.(\ref{R(x)}) is large
\begin{equation}
	|R|\,\gg \, 1~.
	\label{Rinf}
\end{equation} 
Consequently, from Eq.(\ref{dR}) one finds
\begin{equation}
	d\lambda\,\simeq\,\sqrt{z}~\frac{dR}{R}~.
	\label{dR/R}
\end{equation}
From Eq.(\ref{Ai2}) one derives then
\begin{equation}
	{A}_i\,=\,\frac{1}{2\pi i}\,\oint_{C_i}d\lambda \simeq
	\frac{a_i}{2\pi i}\,\oint\,\frac{dR}{R}=a_i~.
	\label{cy}
\end{equation}
Here the cycle $C_i$ is defined in Fig. \ref{FIG1}, the factor $\sqrt{z}$ in the integrand of Eq. (\ref{Ai2}) is approximated by $a_i$ due to Eq.(\ref{zpmapp}), and it is taken into account that circling around the path $C_i$ forces $R$ to rotate by an angle $2\pi$ around the origin counter clock-wise. Equation (\ref{cy}) shows that for weak coupling the vector $a$ defined by Eq.(\ref{ome}), correctly represents the field $A$ and therefore can be considered as the classical approximation for the scalar field. Equation (\ref{cy}) supports the form for $X(z)$ in (\ref{X}).

Similarly, from Eq.(\ref{ADi2}) one derives
\begin{equation}
{A}_{\text{\it D},\,i}\,\simeq\,\frac{1}{2\pi i}\,\oint_{C_{\text{\it D},\,i}} d\lambda \simeq
	\frac{\,a_i}{2\pi i}\,(-2)\,\int_{R_{i,\,\text{min}}}^{R_{i,\,\text{max}}} \,\frac{dR}{R}
	\,\simeq\,
	\frac{i}{2\pi}\, a_i\ln\,\frac{R_{i,\,\text{max}}^{\,2}}{R_{i,\,\text{min}}^{\,2}}~.
	\label{cyD}
\end{equation}
The factor $(-2)$ in the second identity here accounts for the contributions from the two parts of the integration cycle, which run on the first and second sheets of the Riemann surface, as well as the fact that $Y(z)\simeq -R(z)$ on the first sheet in the region of interest. The integration limits in Eq. (\ref{cyD}) can be estimated as follows
\begin{align}
&R_{{i,\,\text{min}}}\,\approx \,1 ~,
\label{Rm1}
\\
&R_{{i,\,\text{max}}}\,\approx \,{(a^2)^{\hv/2}}/{\Lambda^{\hv}}~.	
	\label{Rm}
\end{align} 
The first identity here uses the obvious fact  that $R(z_{i,-})=R(z_{i,+})=1$, where $z_{i,\,\pm}$ are the nodes of $Y(z)$. Consequently,  $R(z)\approx 1$ in some vicinity of the cut stretched between $z_{i,+}$ and $z_{i,-}$. The second equality  is based on the assumption that for all $j=1,\dots r$ an estimate $|a_j^2|\sim |\,[\, a \,]^2|\sim |\,a^2|$ is valid (remember $a^2\equiv a\cdot a$), from which one derives using (\ref{P(z)}) that $P(z)\approx (-1)^r (a^{2})^r$ when $z$ is not close to the cut between $z_{i,+}$ and $z_{i,-}$. 

We can presume that the field $a$ is located in the Weyl chamber, which walls are orthogonal to the simple roots. Then the fact that 
$a_j=a\cdot\av_j$ have all the same order of magnitude means simply that the field $a$ is located well inside the Weyl chamber, being not close to any of its walls. Thus, repeating, Eq.(\ref{Rm}) is valid provided the field $a$ is not close to a wall of the Weyl chamber
(the vicinity of a wall is discussed below in Section \ref{wall}).

Equations (\ref{cyD})-(\ref{Rm}) imply
\begin{equation}
{A}_{\text{\it D},\,i}\,\simeq\,
	\frac{i}{2\pi}\,\hv a_i\ln\,\frac{a^2}{\Lambda^2}~.
	\label{ADLL}
\end{equation}
Substitute now this result together with (\ref{cy}) into (\ref{Aq}),(\ref{ADq}).  Remembering also the definition of the classical field $a$ in Eq. (\ref{ome}) one immediately recovers the fields at weak coupling in Eqs. (\ref{Aweak}) and (\ref{APTh2}).
Importantly, this implies that the coefficient of
the Gell-Mann - Low beta-function is reproduced correctly, $b^{(\text{w})}=2\hv$. The latter fact justifies relation $Q\propto \Lambda^{\hv}$, which was used to define $Q$  in Eq.(\ref{Q(z)}). 

Summarizing, it is verified that Eqs. (\ref{YY}) comply with the weak coupling limit.

\subsection{Scalar field near a wall of Weyl chamber}
\label{wall}

Consider the weak coupling limit in a particular case when one of the coefficient $a_k$ in the expansion of the classical field $a$ in Eq.(\ref{ome}), 
is smaller than others, but still is large in relation to $\Lambda$, 
\begin{equation}
 \Lambda^2\ll |\,a_k|^2\ll |\,a_i|^2\approx |\,a^2|~.
\label{wal}
\end{equation}
Here $i\ne k$, where $k$ is fixed. From a geometrical point of view condition (\ref{wal}) states that the scalar field is chosen sufficiently close to the wall of the Weyl chamber, which is orthogonal to the simple root $\alpha_k$. Using Eq.(\ref{P(z)}) we find then
\begin{equation}
R_{i,\,\text{max}}\,\approx\,
\frac{1}{\Lambda^{\hv}}\times
\left\{\begin{array}{ll} 
(a^2)^{\hv/2}~,           				\quad & i\ne k
\\
{a_k^2\,(a^2)^{\hv/2-\,1}}~,          \quad & i=k 
\end{array}
\right.
\label{Rij}
\end{equation}
which should be used instead of Eq.(\ref{Rm}) when $a_k$ is small.
Substituting it into (\ref{cyD}) we derive
\begin{equation}
{A}_{\text{\it D},\,i}\,\simeq\,
	\frac{i}{2\pi}\,a_i\,\left(\,
	\hv \ln\,\frac{a^2}{\Lambda^2}+2\,\delta_{ik}\,\ln\,\frac{a^2_k}{a^2}~\right)~.
\label{ADlog}
\end{equation}
The first term in the brackets here complies with Eq.(\ref{ADLL}), the second gives the correction, which acknowledges the fact that $a_k$ is small. 
The arguments of the logarithmic function
in these two terms differ. 
The first has a factor $|a^2|\gg \Lambda^2$, which is conventional in the perturbation theory. The second is derived presuming that $|(\alpha_k \cdot A)|^2\ll |A^2|$.
Using (\ref{ADlog}) as well as (\ref{cy}), (\ref{ADq})  one presents the dual field as follows
\begin{equation}
A_\text{\it D}\,\simeq\,\frac{i}{2\pi}\,\left(\hv A\, \ln\,\frac{A^2}{\Lambda^2}+2\,A_k\,\ln\Big(\,\frac{A^2_k}{A^2}\,\Big)~\omega_k\,\right)~.
\label{ADA+log}
\end{equation}
Keeping in mind that the simple coroots and fundamental weights are orthonormal,
one derives from Eq.(\ref{Aq}) that $\ln(\alpha_k\cdot A)^2
\approx \ln(A_k^2)$, where the numerical constant $2\ln\big(2/\alpha_k^2\big)$ was neglected.
Consequently, Eq.(\ref{ADA+log}) can be presented as follows
\begin{equation}
A_\text{\it D}\,\simeq\,\frac{i}{2\pi}\,\left(\,\hv A\, \ln\,\frac{A^2}{\Lambda^2}+
2\,(\av\cdot A)
\,\ln\Big(\frac{(\alpha\cdot A)^2}{A^2}\Big)~\omega_\alpha\,\right)~.
\label{ADA+log+final}
\end{equation}
This result is valid provided $A$ is located near the wall of the Weyl chamber defined by the root $\alpha$. We presumed previously that $\alpha$ is a simple root, but from the derivation it is clear that the result is applicable for any root. 
In line with this fact the notation is Eq.(\ref{ADA+log+final}) is modified, a root $\alpha$ and the corresponding weight $\omega_\alpha$, are used there instead of the previously employed simple root $\alpha_k$, and the fundamental weight $\omega_k$; $\alpha$ and $\omega_\alpha$ can differ from $\alpha_k$ and $\omega_k$ by a Weyl reflection, a similar notation is used below.

%The first term in Eq.(\ref{ADA+log+final}) complies with Eq.(\ref{ADLL}) from the previous Section \ref{inside Weyl chamber}, which in turn agrees with Eq.(\ref{APTh2}) predicted by the perturbation theory. The second term in the brackets in (\ref{ADA+log+final}) is a correction related to the fact that the field $A$ is located close to the wall of the Weyl chamber.  
%The logarithmic function in this correction complies with the predictions of the perturbation theory, which is used directly (without references to the low-energy solution), see Eq.(\ref{ADA+log+PTh}). However, there is a difference in the coefficient in front of this correction. This coefficient is described by the tensor product $2\omega_\alpha\!\otimes \alpha$ in (\ref{ADA+log+final}). The direct application of the perturbation theory
%leads to the different tensor product $\alpha\otimes \alpha$ in (\ref{ADA+log+PTh}).
%Clearly, for weak coupling the perturbation theory applied directly has to be valid. On the other hand, Eq. (\ref{ADA+log+final}) is derived from the proposed model, which complies with a number of verifications, some of them were mentioned above and several more are discussed below. This fact makes the model robust. For the time being let us put  the found discrepancy aside, proceed with the study of the proposed model, and then return to the mentioned difference in Section \ref{Discussion} with more evidence at hand.

\section{Discrete transformations, chiral symmetry and duality}
\label{Chiral+symmetry}
%Consider chiral transformations of the $\CN=2$ supersymmetric gauge theory. On the classical level they manifest themselves via the continuous transformations of the fields
%\begin{align}
%	 \vartheta \rightarrow e^{i\gamma} \vartheta~,\quad
%	\psi  \rightarrow e^{i\gamma} \psi~, \quad
%	\lambda \rightarrow e^{i\gamma} \lambda~,\quad A \rightarrow e^{2i\gamma} A~.
%		\label{2gammaA}
%\end{align}
%Here $\vartheta$ is a conventional anti-commuting variable of the $\CN=1$ superspace \cite{Wess:1991}. Quantum corrections break this symmetry to $Z_{\,4\hv}$; the phase $\gamma$ can take only discrete values
%\begin{equation}
%	\gamma=2\pi\,\frac {m} {4\hv}~,\quad m=0,1,\,\dots \,4 \hv-1~.
%	\label{gamma}
%\end{equation}
%The effect is related to the variation of the $\theta$-angle of the theory due to the chiral transformation, which reads
%%\begin{equation}
%$\Delta \,\theta \,=\,4\hv\gamma$, and which 
%%\label{theta}
%%\end{equation}
%%The factor $4\hv$ here equals the total number of fermion zero-modes in the field of an instanton, see e. g. \cite{Shifman:1999mv}. 
%%In this field each fermion field has $2\hv$ zero modes. Consequently, the two fermion fields $\psi$ and $\lambda$ available in the $\CN=2$ gauge theory make the total number of zero modes $4\hv$. 
%should be an integer of $2\pi$.
%
%
%
Following Eqs.(\ref{2gammaA})  consider the chiral transformation of the classical scalar field
\begin{equation}
a\,\rightarrow \,\exp(\,i\pi/\hv)\,a, 
\label{aexpa}
\end{equation}
where it suffices to take the value $m=1$ in (\ref{gamma}). 
It is convenient here to use the curve and the differential in the scaled notation 
of Eqs.(\ref{zeta})-(\ref{p}), in which Eqs.(\ref{YYY}),(\ref{Ypm}) read
\begin{align}
&\,y\,(\zeta)~~=~\,y_+(\zeta)~y_-(\zeta)~,
\label{yyy}
\\
&y_{\,\pm}\,(\zeta)\,=\,p\,(\zeta)\,\pm \, \Lambda^{\hv}\! /\, [\, a\,] ^{\hv}~.
\label{ypm}
\end{align}
The nodes $\zeta_{\,i,\,\pm}$ of $y_{\,\pm}\,(\zeta)$, which satisfy 
\begin{equation}
y_{\,\pm}\,(\zeta_{\,i,\,\pm})\,=\,0~,
\label{nodes}
\end{equation}
are related to the previously introduced nodes $z_{\,i,\,\pm}$ of $Y_{\,\pm}\,(z)$, see Eq.(\ref{zpm}),  $\zeta_{i,\pm}=z_{\,i,\pm} / [\,a\,]^2$.
 
Observe that $\xi_i=a_i/[ a\,]$ from Eq.(\ref{zeta}) are invariant under the chiral transformation 
(\ref{aexpa}), which allows one to presume that  $\zeta$ is also invariant under this transformation. Precisely this  property makes this variable more convenient than $z$. Observe further that 
under the transformation (\ref{aexpa}) the second term in Eq.(\ref{ypm}) changes its sign. This means that this chiral transformation induces 
\begin{equation}
y_{\,\pm}\,(\zeta)\,\rightarrow\,y_{\,\mp}\,(\zeta)\,,
\label{ypmmp}
\end{equation}
which keeps $y(\zeta)$ invariant while enforces the nodes to
swap their places
\begin{equation}
\zeta_{\,i,\,+}\,\rightarrow \,\zeta_{\,i,\,+}^{\,\prime}\,\equiv\,\zeta_{\,i,\,-} ~,\quad\quad
\zeta_{\,i,\,-}\,\rightarrow\,\zeta_{\,i,\,-}^{\,\prime} \,\equiv\,
\zeta_{\,i,\,+}~.
\label{zetapm}
\end{equation}
It is instructive to relate the chiral transformation (\ref{aexpa}) to a continuous transformation $a\rightarrow \exp(i\gamma)a$  with a continuous variable $\gamma$, 
which is allowed to run over the interval $0\le \gamma\le \pi/\hv$. One can then follow the variation of the nodes $\zeta_{\,i\,\pm}$ under this continuous transformation finding  that 
when $\gamma$ varies from $0$ to $\pi/\hv$ the points $\zeta_{\,i,\,+}$ and  
$\zeta_{\,i,\,-}$ exhibit rotation over an angle $\pi$ counter clockwise around their mutual center. To verify this claim take the weak coupling region $|a^2|\gg \Lambda^2$ discussed in Section \ref{weakcoupling}, where the proof is straightforward.
\begin{figure}[tbh]
  \centering \includegraphics[ height=7cm, keepaspectratio=true]{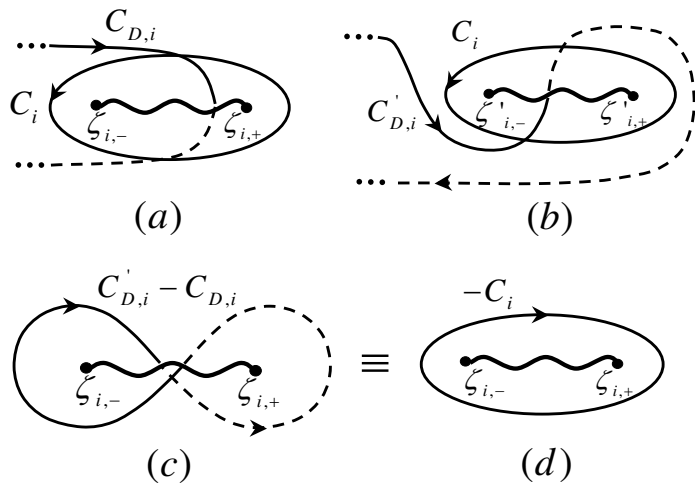}
\vspace{0cm}
\caption{Chiral transformation for the cycles $C_i$ and $C_{\text{\it D},\,i}$. Solid and dashed lines show behavior of cycles in the vicinity of $\zeta_{\,i,\,\pm}$ on the first and second sheets of the Riemann surface respectively, wavy line indicates the cut, (a) shows the initial location of the cycles, compare Fig. \ref{FIG1} , (b) the result of the chiral transformation (\ref{aexpa}),(\ref{zetapm}), which produces a variation of $C_{\text{\it D},\,i}\rightarrow C_{\text{\it D},\,i}^{\,\prime}$, (c) the difference of $C_{\text{\it D},i}^{\,\prime}-C_{\text{\it D},i}$,
(d) same  difference presented via $-C_i$. }\label{FIG2} 
   \end{figure}
   \noindent

Adopting a picture, in which the  chiral transformation amounts to rotation of the 
nodes $\zeta_{\,i\,\pm}$ by the angle $\pi$, we find the influence of the chiral transformation (\ref{aexpa}) on the integration cycles. Fig. \ref{FIG2} shows behavior of the cycles $C_i$ and $C_{\text{\it D},\,i}$ in the vicinity of the nodes $\zeta_{i,\,\pm}$. Fig. \ref{FIG2} (a) is similar to Fig.  \ref{FIG1}, though is presented using the scaled parameters $\zeta_{\,i,\,\pm}$ (instead of $z_{\,i,\,\pm}$ employed in Fig. \ref{FIG1}). Fig. \ref{FIG2} (b) shows that the chiral transformation of the nodes described by Eq.(\ref{zetapm}) induces a variation of the dual integration cycle, $C_{\text{\it D},\,i}\rightarrow C_{\text{\it D},\,i}^{\,\prime}\,$, while keeping $C_{i}$ intact. Figures (c) and (d) show that this variation amounts to
\begin{equation}
C_{\text{\it D},\,i}\,\rightarrow \,C_{\text{\it D},\,i}^{\,\prime}\,\equiv \,C_{\text{\it D},\,i}-C_{i}~.
\label{CCC}
\end{equation}
The equality here means that the integration of the differential $d\lambda$ over the cycles on the left and right hand sides of the identity produces the same result. It is instructive to present Eq.(\ref{CCC}) in the matrix notation. Introduce the $2r$ vector of cycles
\begin{equation}
\Sigma\,=\,\big(\,C_{\text{\it D},\,1},\dots C_{\text{\it D},\,r},\,C_1,\dots C_r \,\big)^T~.
\label{3C}
\end{equation}
Then Eq.(\ref{CCC}) and the fact that $C_i$ remains invariant under the chiral transformation mean that this transformation for the cycles reads
\begin{equation}
\Sigma\,\rightarrow \,\Sigma\,^\prime\,=\, H \,\Sigma~,
\label{CMC}
\end{equation}
where the $2r\times 2r$ matrix $H$ is defined in Eq.(\ref{Mch}).
Note a similarity 
between Eq.(\ref{CMC}) and the chiral transformation for the fields in Eq. (\ref{ADgamma}).

Using Eqs. (\ref{scale}) and (\ref{CCC}) one finds the chiral transformation for
$A_{\,i}$ and $A_{\text{\it D},\,i}$ defined in Eqs.(\ref{Ai2}),(\ref{ADi2})
\begin{align}
&\,A_{\,i}~~\rightarrow ~A_{\,i}^{\,\prime}~~\,=\,\exp\,(\,i\,\pi/\hv\,)\,A_{\,i} ~.
\label{AA}
\\
&A_{\text{\it D},\,i}\rightarrow \,A_{\text{\it D},\,i}^{\,\prime}\,=\,\exp\,(\,i\,\pi/\hv\,)\,\big(\,A_{\text{\it D},\,i}-A_{\,i}\big) ~.
\label{AAA}
\end{align}
The factor $\exp\,(\,i \,\pi/\hv\,)$ here arises from the factor $[\, a\,]$ in Eq.(\ref{scale}), while the two terms in the brackets in (\ref{AAA}) come from the two terms in (\ref{CCC}). Combining Eqs. (\ref{AA}),(\ref{AAA}) with the definitions of the fields in Eqs.(\ref{Aq}),(\ref{ADq}) we see that the fields are transformed in full accord with Eq.(\ref{ADgamma}). Thus the model  proposed  reproduces the chiral symmetry, as it should.

Consider now another important discrete transformation, duality. Observe that the invariance under the duality transformation is incorporated in the theory from the very beginning. To see this point take Fig. \ref{FIG1}, which illustrates the Riemann surface. One immediately derives from this figure that the transformation of the cycles
\begin{align}
&C_i~~~\rightarrow ~C_{i}^{\,\prime}~~\,=\,-C_{\text{\it D},\,i}~,
\label{CC'}
\\
&C_{\text{\it D},\,i}\,\rightarrow\, C_{\text{\it D},\,i}^{\,\prime}\,=~~\,C_i
\label{CDCD'}
\end{align}
leaves intact the canonical interception form (\ref{CC}),(\ref{CC=d}). Obviously,  it also does not change the differential (\ref{YY}). Therefore 
applying Eqs.(\ref{CC'}),(\ref{CDCD'}) to the scalar fields, which are defined  in Eqs.(\ref{Aq}),(\ref{ADq}), one immediately verifies that the fields satisfy the duality condition (\ref{dual}), as they should.

\section{Strong coupling}
\label{Strong-coupling} 
In the strong coupling limit the masses of dyons are small \cite{Seiberg:1994rs}. According to  Eqs. (\ref{MZ}),(\ref{Z}) this means that the fields $A,A_\text{\it D}$ satisfy
\begin{equation}
%\CZ_{\,\CG}\,=\,\CG \,\varPhi\,=\, 
\CG\,\Phi\,=\,g \cdot \AD+q\cdot A\,\rightarrow \, 0~,
\label{m0} 
\end{equation}
where $\CG=(g,q)$ is the dyon charge. Take the simplest case, the monopole. According to Eq.(\ref{ma})  its charge equals $\CG_{\,\alpha_i, \,0}=(\av_i,0)$, where $\av_i$ is a simple root. In this case Eq.(\ref{m0}) shows that the strong coupling limit takes place when
\begin{equation}
\CG_{\,\alpha,\,0}\,\Phi\,=\,\av_i\cdot A_\text{\it D}\,=\,A_{\text{\it D},\,i}\,\rightarrow\, 0~.
\label{ADi->0}
\end{equation}
The identity here takes into account the expansion of the field over the set of fundamental weights in Eq.(\ref{ADq}). Note that an important statement, which asserts that at strong coupling the monopole is light, is expressed in Eq.(\ref{ADi->0}) in a simple, appealing form. At this point an advantage of the basis of fundamental weights becomes evident. The orthonormal condition between the simple roots and fundamental weights (\ref{avo})
%\begin{equation}
%\av_i\cdot\omega_j\,=\,\delta_{ij}~, 
%\label{avo}
%\end{equation}
guarantees that second identity in Eq.(\ref{ADi->0}) holds, and therefore discussing the massless monopole we need to focus our attention on one and only one term $A_{\text{\it D},\,i}\,\omega_i$ in the expansion of the field $A_\text{\it D}$ in Eq.(\ref{ADq}). There is also a simple, attractive geometrical implication of Eq.(\ref{ADi->0}). Condition $\av_i\cdot A_{\text{\it D}}=0$ means that 
%\begin{equation}
%A_{\text{\it D}}\cdot \alpha_i=0~.
%\label{Aal}
%\end{equation}
%Henceforth, 
the monopole becomes massless when the dual scalar field $A_\text{\it D}$ hits a wall of the Weyl chamber. 
%Note that the basis of the fundamental weights, which was used in Eqs.(\ref{ome}), (\ref{Aq}) and (\ref{ADq}) for the scalar field and its dual, allows one to express this important condition in a very simple form.

Figure \ref{FIG1} shows that  in order to satisfy Eq.(\ref{ADi->0}) it is necessary to bring either $z_{i,\,-}$ or $z_{i,\,+}$ to the origin. Suppose it is $z_{i,\,-}$, i.e. suppose that
\begin{equation}
z_{i,\,-}\rightarrow 0~. 
\label{z0}
\end{equation}
(The case of $z_{i,\,+}\rightarrow 0$ reveals similar properties since the chiral symmetry interchanges $z_{i,\,-}$ and $z_{i,\,-}$, see Section \ref{Chiral+symmetry}).  Our next goal is to find the singularity in the fields $A$ and $A_\text{\it D}$ in the limit specified in (\ref{ADi->0}). This can be achieved by using relatively simple analytic approach. For the SU(2) gauge group the necessary calculations were presented in \cite{Seiberg:1994rs}. 
\begin{figure}[tbh]
  \centering \includegraphics[ height= 5cm, keepaspectratio=true]{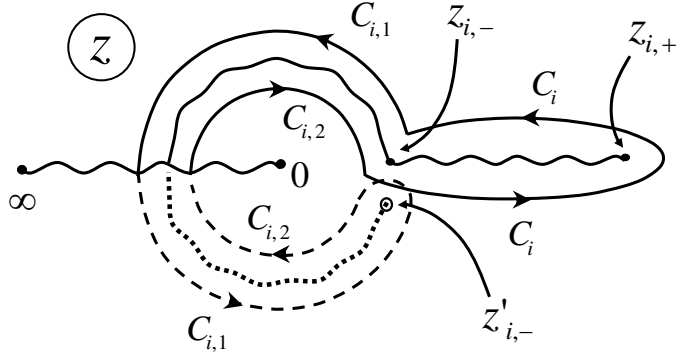}
\vspace{0cm}
\caption{The cycle $C_i$, which connects $z_{i,\,-}$ with $z_{i,\,+}$ and the wavy lines, which represent the two cuts, have the same meaning as in Fig. \ref{FIG1}. Under the rotation of $z_{i,\,-}$ around the origin the cycle $C_i$ acquires an extension described by two segments. One of them,  called $C_{i,\,1}$, runs from $z_{i,\,-}$ around the origin on the first sheet of the Riemann surface and then continues on the second sheet finishing at $z_{i,\,-}^{\,\prime}$. Another segment $C_{i,\,2}$ runs in the opposite direction on the other side of the cut. The slashed lines show parts of these segments on the second sheet. The dotted wavy line presents the extension of the cut between $z_{i,\,-}$ and $z_{i,\,+}$, which appears when $z_{i,\,-}$ is brought to its position on the second sheet $z_{i,\,-}^{\,\prime}$.
}\label{FIG3} 
   \end{figure}
   \noindent
However, in general case more appealing looks an approach based on a topological argument, which we pursue below using Fig. \ref{FIG3} for illustration. In accord with Eq.(\ref{ADi->0})  we bring  $z_{i,\,-}$ close to the origin, which can be achieved by tuning the parameters $a_i$ in the curve $Y(z)$ defined by Eq.(\ref{YY}). Then we rotate $z_{i,\,-}$, say, counterclockwise around the origin by an angle $2\pi$ (again by tuning appropriately the parameters $a_i$), bringing it in the end to the position called $z_{i\,-}^{\,\prime}$ in Fig. \ref{FIG3}, and stretching appropriately the cut attached to it. Observe that this operation does not affect any $C_{\text{\it D},\,j}$, $j=1,\dots r$. Hence the periods related to these cycles remain the same. This makes the field $A_\text{\it D}$, which is expressed in terms of these periods, invariant under the rotation of $z_{i,\,-}$
\begin{equation}
A_\text{\it D}\,\rightarrow \,A^{\,\prime}_\text{\it D}\,=\,A_\text{\it D}~. 
\label{ADinv}
\end{equation}
In contrast, the rotation of $z_{i,\,-}$ stretches the cycle $C_i$, which acquires an additional part. For convenience this part is divided in Fig. \ref{FIG3} into 
two segments called  $C_{i,\,1}$ and $C_{i,\,2}$. The first one starts from $z_{i,\,-}$ on the first sheet of the Riemann surface and circles around the origin over an angle of $2\pi$. The segment $C_{i,\,2}$ starts from $z_{i,\,-}^{\,\prime}$ on the second sheet of the Riemann surface and makes the journey in the opposite direction by rotating around the origin clockwise being located on the other side of the cut (the cut that runs between $z_{i,\,-}^{\,\prime}$ and $z_{i,\,+}$) from the segment $C_{i,\,2}$. The integrand contains only square root functions. Therefore the values of the integrand on the $C_{i,\,1}$ and $C_{i,\,2}$ are opposite in sign. Since the directions of these two segments are also  opposite, their contributions to the integral are same
$\int_{C_{i,\,1}}d\lambda=\int_{C_{i,\,2}}d\lambda$. Comparing Figs. \ref{FIG1} and \ref{FIG2} we observe also that  $C_{i,\,1}=-C_{\text{\it D},\,i}$ concluding from this that the integration over the part of the cycle $C_i$, which is stretched  when $z_{i,\,-}$ is rotated around the origin, equals
\begin{equation}
\frac{1}{2\pi i}\int_{C_{i,1}+C_{i,2}}\!\!\!\!d\lambda\,=\,
-2\,\frac{1}{2\pi i}\int_{C_{\text{\it D},\,i}}\!d\lambda\,=\,-2A_{\text{\it D},\,i}~.
\label{same}
\end{equation}
We deduce from this that the rotation of $z_{i,\,-}$ around the origin results in the variation of the component $A_i$ of the scalar field
\begin{equation}
A_i\,\rightarrow\,A_i^{\,\prime}\,=\,\frac{1}{2\pi i}\int_{C_i+C_{i,1}+C_{i,2}}\!\!\!\!d\lambda\,=\,A_i-2A_{\text{\it D}_,\,i}~.
\label{Ai->}
\end{equation}
All other components $A_j$, $j\ne i$, are kept intact. Therefore the transformation (\ref{Ai->}) for the scalar field can be conveniently rewritten as follows
\begin{equation}
A\,\rightarrow \,A^{\,\prime}\,=\,A-2\omega_i \,(\av_i\cdot A)~. 
\label{Arot}
\end{equation}
Combining this with Eq.(\ref{ADinv}) we write the transformation of the field $\Phi$  due to rotation of $z_{i,\,-}$ around the origin 
\begin{equation}
\Phi\,\rightarrow \Phi^{\,\prime}\,=\,M_\alpha\,\Phi~,
\label{2pi}
\end{equation}
where $M_\alpha$ is the following monodromy 
\begin{equation}
M_\alpha\,=\,\begin{pmatrix} \quad 1 & 0\\-2\,\omega_\alpha\!\otimes\av &1 \end{pmatrix}~.
\label{Malpha}
\end{equation}
Here the subscript $i$ for the root and weight are omitted since it is clear from the derivation that the result is applicable to any root $\alpha$ and the corresponding weight $\omega_\alpha$, which can be transformed by Weyl reflections into the pair of a simple root and the corresponding fundamental weight.

The monodromy $M_\alpha$ in Eq.(\ref{2pi}) was introduced via the rotation of the node of $z_{i,\,-}$ around the origin. It is essential that this rotation was considered when this root is close to the origin. This ensures that no other topological properties of the Riemann surface interfere with this transformation. At the same time this closeness indicates, as we know from Eq.(\ref{m0}), that the mass of the monopole turns zero. 
We conclude from this observation that the model allows the monopole with the charge $g=\av$ to become massless. In addition, we learn that the monodromy $M_\alpha$ exists because this massless monopole is present in the theory.

To make Eq.(\ref{2pi}) more transparent it is convenient to split each of the fields $A$ and $A_\text{\it D}$ into the two components presenting them as follows
\begin{align}
&A~~=~\omega_\alpha \,(\,\av\cdot A\,)~+
(1-\omega_\alpha \otimes \av)\,A~,
\label{splitA}
\\
&A_\text{\it D}\,=\,\omega_\alpha \,(\av\cdot A_\text{\it D})+
(1-\omega_\alpha \otimes \av)\,A_\text{\it D}~.
\label{splitAD}
\end{align}
The first terms here are aligned along the weight $\omega_\alpha$, while the second ones are orthogonal to the root $\alpha$, and therefore belong to the wall of the Weyl chamber defined by this root. Clearly, the variation of $\Phi$ described by Eqs.(\ref{2pi}) and (\ref{Malpha}) takes place only for the components of the fields that point along the weight $\omega_\alpha$, i.e. the first terms in  (\ref{splitA}), (\ref{splitAD}).  It is convenient therefore to introduce a 2-dimensional vector of these components of the fields
\begin{equation}
\Phi^{(2)}_\alpha\,=\,\begin{pmatrix} ~\av\cdot A_\text{\it D}\\
\av\cdot A \end{pmatrix}~.
\label{PhiTan}
\end{equation} 
and rewrite Eq.(\ref{2pi}) for these components
\begin{align}
&\Phi^{(2)}_{\,\alpha}\,\rightarrow\,\Phi^{(2)\,\prime}_{\,\alpha}\,=\,
M^{\,2\times 2}_{\,\alpha}\,\Phi^{(2)}_{\,\alpha}~,
\label{P2}
\\
&M^{\,2\times 2}_{\,\alpha}\,=\,\begin{pmatrix} ~~ 1 & 0~\\-2&1 ~\end{pmatrix}~.
\label{M22}
\end{align}
As was mentioned, the components of the fields $A_{\text{\it D}}$ and $A$ that belong to the wall of the Weyl chamber defined by the root $\alpha$ are not transformed in Eq.(\ref{2pi}). Hence a $2\times 2$ matrix $M^{\,2\times 2}_{\,\alpha}$ represents all essential properties of the monodromy $M_\alpha$. 

Compare now $ M^{\,2\times 2}_{\,\alpha}$ with the monodromy, which describes the monopole in the Seiberg-Witten solution for the theory with the SU(2) gauge group. The latter was  called in \cite{Seiberg:1994rs}  $M_1^\text{SU(2)}$ (the superscript SU(2) for $M_1$, as well as for several other operators from the Seiberg-Witten solution discussed below is added here). Interestingly, the two monodromies are identical
\begin{equation}
M^{\,2\times 2}_{\,\alpha}\,=\,M_1^\text{SU(2)}~.
\label{M=M}
\end{equation} 
This shows that properties of the monopole in the $\CN=2$ supersymmetric theory with an arbitrary gauge group have close correspondence with the monopole in the theory based based on the SU(2) group. This is not entirely surprising, though the transparency of Eq.(\ref{M=M}) is rewarding.

It is instructive now to return to the component presentation of the monodromy in Eq.(\ref{Ai->}). One can consider it as a statement that $A$ exhibits the logarithmic singularity at $z_{i,-}\rightarrow 0$. In contrast, $A_\text{\it D}$ is regular in this limit and satisfies
$A_{\text{\it D},\,i}\rightarrow 0$. 
From Eq. (\ref{Ai->}) one finds the coefficient, which governs the logarithmic singularity in $A$, obtaining
\begin{equation}
A_j\,\approx\,A_j^{(0)}+\frac{i}{2\pi}\,A_{\text{\it D},\,i}\,\ln\left(\,\frac{A_{\text{\it D},\,i}^2}{\Lambda^2}\,\right)~\delta_{ji}~,
\label{Alog}
\end{equation} 
where the superscript $(0)$  refers to the nonsingular term.  
An alternative way to derive this result gives Eq.(\ref{M=M}). Using it, one can argue that Eq.(\ref{Alog}) is valid for any gauge group since it is applicable for the case of the SU(2) group.

From Eq.(\ref{Alog}) one finds that the logarithmic singularity is present in the dual matrix of coupling constants 
\begin{equation}
\big(\tau_{\text{\it D}}\big)_{j\,k}\,=\,-\frac{\partial A_j}{\partial A_{\text{\it D},\,k}}\,\approx\,\big(\,\tau_{\text{\it D}}^{(0)}\,\big)_{j\,k}-
\frac{i}{2\pi}\,\ln\left(\,\frac{A_{\text{\it D},\,i}^2}{\Lambda^2}\right)\,\delta_{ji}\,\delta_{ki}~.
\label{tauDlog}
\end{equation}
The logarithmic term makes the diagonal matrix element with $j=k=i$ large, which ensures that the matrix $\tau_{\text{\it D}}$  possesses a large eigenvalue
$\tau_{\text{\it D},\,\text {eig}}\approx(-i/2\pi)\,\ln(\,A_{\text{\it D},\,i}^2/\Lambda^2\,)$, as it should. Correspondingly, Eq.(\ref{tauDlog}) also reproduces the necessary coefficient $b^\text{(s)}=-2$ of the Gell-Mann - Low beta-function at strong coupling.
%Another important outcome of this discussion is that the monopole with the charge $\CG_{\,\alpha_i,\,0}=(\av_i,0)$, which is specified by a simple coroot $\av_i$, can become massless, in agreement with predictions of \cite{Kuchiev:2008mv}. 

Consider now the case of dyons proper. According to Eq.(\ref{ma}) the charges of dyons are $\CG_{\,\alpha_i,\,m}=(\av_i,-m\av_i)$.
Let us  describe this case by modifying the proposed above description of monopoles. The model we discuss complies with the discrete chiral symmetry, see Section \ref{Chiral+symmetry}. Eqs.(\ref{ADgamma}),(\ref{Mch}) show how the field $\Phi$ is affected by chiral transformations. Comparing this result with Eqs.(\ref{MZ}),(\ref{Z}), which describe the mass of a dyon, one concludes 
that the chiral transformation of the field $\Phi$ induces the following chiral transformation of the dyon charge, compare \cite{Kuchiev:2008mv},
\begin{equation}
\CG\,\rightarrow \,\CG^{\,\prime}\,=\,\CG\,H~,
\label{GG'}
\end{equation}
where $H$ is from Eq.(\ref{Mch}). Thus, a chiral transformation applied to the monopole charge $\CG=\CG_{\,\alpha_i,\,0}$ results in the dyon with the charge $\CG^{\,\prime}=\CG_{\,\alpha_i,1}$. Applying this procedure $m$ times one transforms the monopole into  the dyon 
\begin{equation}
\CG_{\,\alpha_i,\,m}\,=\,\CG_{\,\alpha_i,\,0}\,H^m~.
\label{GGm}
\end{equation}

Consider now Eq.(\ref{m0}) for the dyon taking there $\CG=\CG_{\,\alpha,\,m}$, writing this equation in a form $\CG_{\alpha_i,\,m}\,\Phi \rightarrow 0$, and aiming to derive properties of $\Phi$ from the latter condition. Equation (\ref{GGm}) allows one to rewrite this as $\CG_{\,\alpha_i,\,0}\,M^m\,\Phi\rightarrow 0$.
We see from here that $M^m\,\Phi$ possesses properties, which are similar to
$\Phi_\text{mon}$, where $\Phi_\text{mon}$ is the scalar field, which was previously studied for the case of the monopole (and which satisfies $\CG_{\,\alpha_i,\,0}\,\Phi_\text{mon}\rightarrow 0$).
Thus, we can state a relation between the field $\Phi$, which satisfies  Eq.(\ref{m0}) for the dyon, and the field $\Phi_\text{mon}$ that satisfies the same condition for the monopole 
\begin{equation}
\exp(\,\pi i m/\hv\,)\,H^m\,\Phi\,=\,\Phi_\text{mon}~.
\label{MmPP}
\end{equation}
Let us repeat, $\Phi_\text{mon}$ describes the light monopole. It is the field, which satisfies condition $A_{\text{\it D},\,i}=0$ from Eq.(\ref{ADi->0}).  Hence, using this field, one constructs the field for the light dyon from Eq.(\ref{MmPP}), thus satisfying (\ref{m0}) for the dyon. This implies that the model considered describes light dyons with the charges $\CG_{\,\alpha,\,m}$, as it should according to \cite{Kuchiev:2008mv}.

Eq. (\ref{MmPP}) also shows how to construct the monodromy, call it $M_{\,\alpha,\,m}$,
which corresponds to the dyon and results in the following transformation of the dual field
\begin{equation}
\Phi\rightarrow \Phi^{\,\prime}=M_{\,\alpha,\,m}\,\Phi~. 
\label{dyon}
\end{equation}
From Eq.(\ref{MmPP}) we expect that $H^m\,\Phi^{\,\prime}=\Phi^{\,\prime}_\text{mon}$, where 
$\Phi^{\,\prime}_\text{mon}$ is the monodromy for the monopole. 
The latter, according to Eq.(\ref{2pi}), reads $\Phi^{\,\prime}_\text{mon}=M_\alpha\Phi_{\text{mon}}$. Using Eq.(\ref{Malpha}) for $M_\alpha$ we find the monodromy for the dyon in Eq.(\ref{dyon}) 
\begin{equation}
M_{\,\alpha,\,m}\,=\,H^{-m}\,M_{\,\alpha}\,H^{m}\,=\,
\begin{pmatrix} 1-2\,m \,\omega_\alpha\!\otimes \av & 2\,m^2\, \omega_\alpha\!\otimes \av \\
-2 \,\omega_\alpha\!\otimes \av                  & 1+2 \,m\,\omega_\alpha\!\otimes \av  
\end{pmatrix}~.
\label{Mm}
\end{equation}
To make this result more transparent let us use the 2-dimensional vector $\Phi^{(2)}_\alpha$ introduced in Eqs.(\ref{PhiTan}).
Equations (\ref{dyon}),(\ref{Mm}) in this notation read
\begin{align}
&\Phi^{(2)}_{\,\alpha}\,\rightarrow \,\Phi^{(2)\,\prime}_{\,\alpha}\,=\,
M^{\,2\times 2}_{\,\alpha,\,m}~\,\Phi^{(2)}_{\,\alpha}~,
\label{Phim2}
\\
&M^{\,2\times 2}_{\,\alpha,\,m}\,=\,\begin{pmatrix}1-2m & 2m^2\\-2 &1+2m \end{pmatrix}~,
\label{Malpham22}
\end{align}
where the $2\times 2$ matrix $M^{\,2\times 2}_{\,\alpha,\,m}$ represents the essential properties of $M_{\,\alpha,\,m}$.
It is instructive to compare this result with the Seiberg-Witten solution for the SU(2) gauge group \cite{Seiberg:1994rs}, where the monodromy for the dyon was called $M_{-1}^\text{SU(2)}$. Taking $m=1$ in Eq. (\ref{Malpham22}), we find
\begin{equation}
M^{\,2\times 2}_{\,\alpha,\,1}\,=\,
\begin{pmatrix}
-1 & 2~ \\
-2 &3~
\end{pmatrix}\,=\,
M_{-1}^\text{SU(2)}~,
\label{M1M-1}
\end{equation} 
which again, remember Eq.(\ref{M=M}), reminds one that properties of the theory at strong coupling for an arbitrary gauge group are closely associated with the theory guided by the SU(2) gauge group.

The associated with the dyon monodromy $M_{\alpha,\,m}$ in Eqs.(\ref{dyon}),(\ref{Mm}) can be interpreted as a presence of the logarithmic singularity in the fields $A_\text{\it D},A$, which exists on condition that the linear combination $A_\text{\it D}-mA$ of the fields is 
singular free, and that its $i$-th component is small, $A_{\text{\it D},\,i}-mA_i\rightarrow 0$. From Eq. (\ref{Malpham22}) one finds the coefficient, which governs the singularity, deriving 
\begin{equation}
A_j~~\approx~
 A_j^{(0)}\,+\,(A_{\text{\it D},\,i}-m\,A_i^{(0)})~\frac{i}{2\pi}\ln\Big[\frac{(A_{\text{\it D},\,i}-m\,A_i^{(0)})^2}{\Lambda^2} \Big]
~\,\delta_{ji}~.
 \end{equation}
We see that the dual matrix of coupling constants possesses a large eigenvalue 
$\tau_{\text{\it D}\text,\, {eig}}\approx(-i/2\pi)\,\ln[\,(A_{\text{\it D},\,i}-m A_i)^2/\Lambda^2\,]$, as expected, and that the resulting 
coefficient of the Gell-Mann - Low beta-function complies with the value $b^{(\text{s})}=-2$, compare similar discussion for the monopole after Eq.(\ref{tauDlog}). 

Summarizing, the model proposed describes light dyons with charges $\CG_{\,\alpha,\,m}=(\av,-m\av)$ in accord with predictions of \cite {Kuchiev:2008mv}. The basis of fundamental weights used for the fields $A$ and $A_\text{\it D}$ makes the derivation transparent.

\section{Combining monodromies for weak and strong coupling}
\label{Relation}

\begin{figure}
  \centering \includegraphics[ height=6 cm, keepaspectratio=true]{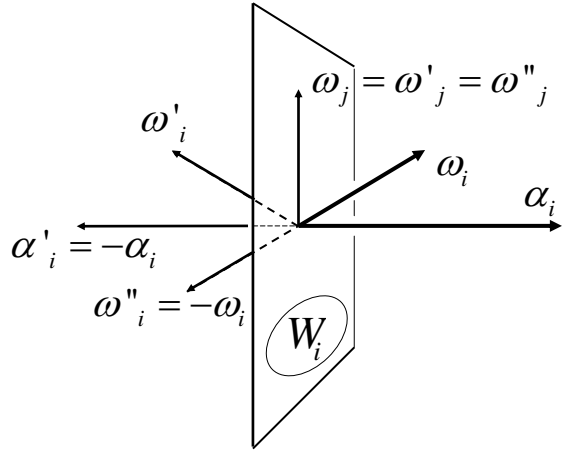}
\vspace{0cm}
\caption{The simple root  $\alpha_i$ defines the wall of the Weyl chamber $W_i$. 
The operator $\sigma_{\alpha_i}$ (\ref{sigma}) keeps the fundamental weights $\omega_j \in W_i$, $j\ne i$, same, though inverses the weight $\omega_i$, $\omega_i\rightarrow
\omega_i^{\,\prime\prime}\equiv\sigma_{\alpha_i}\omega_i=-\omega_i$, see (\ref{Ref-omega}).
The  Weyl reflection keeps the fundamental weights $\omega_j \in W_i$, $j\ne i$ intact, but transforms the simple root $\alpha_i$, $\alpha_i\rightarrow \alpha_i^{\,\prime}$, resulting in the variation of the fundamental weight $\omega_i$, $\omega_i\rightarrow \omega_i^{\,\prime}$.
}\label{FIG4} 
   \end{figure}

Remember Eq.(\ref{dyon}), which describes the monodromy at strong coupling for a dyon with the charge $\CG_\alpha=(\av,-m \av)$. Using Eq.(\ref{Mm}) one finds that the matrix $M_{\,\alpha,\,m}$ of this monodromy satisfies an identity
\begin{equation}
M_{\,\alpha,\,m}\,M_{\,\alpha,\,m+1}\,=\,O_\alpha\,\equiv\,
\begin{pmatrix}
1-2\,\omega_\alpha\!\otimes\av & 2\,\omega_\alpha \!  \otimes\av
\\
0                            &  1-2\,\omega_\alpha\!\otimes\av
\end{pmatrix}~.
\label{MMIW}
\end{equation} 
where the right hand side defines the $2r\times 2r$ matrix $O_\alpha$.
It is interesting that this matrix does not depend on $m$, which appears on the left-hand side. To study other engaging properties of this matrix it is convenient to present it as follow
%\begin{equation}
%O_\alpha\,=\,
%\begin{pmatrix}
%1-2\,\omega_\alpha\!\otimes\av & 2\,\omega_\alpha \!  \otimes\av
%\\
%0                            &  1-2\,\omega\!\otimes\av
%\end{pmatrix}~.
%\label{W}
%\end{equation} 
%It is instructive to present this latter matrix in the form
\begin{equation}
O_\alpha\,=\,I_\alpha\,Q_\alpha\,~,
\label{OIQ1}
\end{equation}
where
\begin{align}
&I_\alpha\,=\,
\begin{pmatrix}
1-2\,\omega_\alpha \!\otimes\av & 0
\\
0                            & 1-2\,\omega_\alpha \!\otimes\av
\end{pmatrix}~,
\\
&Q_\alpha\,=\,
\begin{pmatrix}
~1 &~~ -2\,\omega_\alpha \!\otimes\av~
\\
~0                            & \quad\quad 1~
\end{pmatrix}~.
\label{Q}
\end{align}
The operator associated with the $r\times r$ matrix
\begin{equation}
 \sigma_\alpha=1-2\omega_\alpha \!\otimes\av~, 
\label{sigma}
\end{equation}
which appears in the block-diagonal matrix $I_\alpha$ satisfies condition
\begin{equation}
\sigma_\alpha\,\omega_\beta\,=\,(1-2\delta_{\alpha,\beta})\,\omega_\alpha~,
\label{Ref-omega}
\end{equation}
where $\alpha$ and $\beta$ are two simple roots and $\omega_\alpha,\,\omega_\beta$ are the corresponding fundamental weights. Thus this operator keeps $r-1$ fundamental weights $\omega_\beta$, $\beta\neq \alpha$, intact, while inversing the direction of the weight $\omega_\alpha$. In other words, this operator keeps the wall of the Weyl chamber defined by the root $\alpha$ intact, while producing the  `weight-inversion' on the weight $\omega_\alpha$, $\omega_\alpha\rightarrow \sigma_\alpha \omega_\alpha= -\omega_\alpha$. 
Fig. \ref{FIG4} illustrates this property. 
The operator $\sigma_\alpha$ should not be confused with the Weyl reflection. As was mentioned, $\sigma_\alpha$ produces the weight-inversion for the particular weight specified by a chosen root $\alpha$. In contrast, the Weyl reflection results in the inversion of this root, $\alpha\rightarrow -\alpha$. There are though common features for these two operators. They both keep the wall of the Weyl chamber invariant, and each of them applied to a vector forces this vector to cross the wall of the Weyl chamber, see Fig. \ref{FIG4}.
   \noindent

This discussion shows that the operator $I_\alpha$ fulfills the operation of the weight-inversion for the $2r$-dimensional field $\Phi$. There is a subtlety though.
According to Eq.(\ref{ADA+log+final}) the operation of the weight-inversion
$A\rightarrow A^{\,\prime}=\sigma_\alpha\,A$ for the field $A$
is accompanied by an additional variation of the field $A_\text{\it D}$, which stems from the logarithmic term in this equation. As a result an extra term proportional to $A$ arises in the dual field $A_\text{\it D}$. Observe now that a presence of this term is precisely described by the operator $Q_\alpha$ from Eq.(\ref{Q}). We can assert  therefore that $I_\alpha$ describes the weight-inversion in the classical approximation, while $Q_\alpha$ describes the quantum correction. 

Thus the operator $O_\alpha$ from (\ref{MMIW}) can be considered as a monodromy that describes the weight-inversion combining the classical and quantum parts of the description. Remember now that the monodromies $M_{\alpha,\,m}$ from Eq.(\ref{Mm})
on the heft-hand side of  this equation describe the dyons at strong coupling. In difference, properties of the monodromy $O_\alpha$ on the right-hand side of (\ref{MMIW}) reveal themselves through Eq.(\ref{ADA+log+final}), which is valid for strong scalar field in the region (\ref{wal}), where the perturbation theory is applicable.

There is another interesting interpretation of   Eq.(\ref{MMIW}).
Simple calculation shows that the operator $Q_\alpha$ in Eq.(\ref{Q}), which describes the quantum part of the weight-inversion, is related to the operator $M_\alpha$ in (\ref{Malpha}), which describes the monodromy related to the monopole
\begin{equation}
Q_\alpha\,=\,\Omega\,M^{-1}_\alpha\,\Omega^{-1}~.
\label{OMO}
\end{equation}
Here $\Omega$ is a matrix of  the duality transformation (\ref{Omega}). Consequently, Eq.(\ref{MMIW}) can be rewritten as an identity that expresses the classical part of the weight-inversion $I_\alpha$ via the monodromies $M_\alpha$, $M_{\alpha,\,m}$, $M_{\alpha,\,m+1}$ related to the monopole and two dyons, as well as the duality transformation $\Omega$
\begin{equation}
M_{\,\alpha,\,m}\,M_{\,\alpha,\,m+1}\,\Omega\,M_\alpha\,\Omega^{-1}\,=\,I_\alpha~.
\label{MMM}
\end{equation} 
To simplify description note that when the operator $O_\alpha$ is applied to the field $\Phi$ it affects non-trivially only those components of the scalar and dual fields, which are longitudinal in respect to $\omega_\alpha$, while keeping all other components, which belong to the wall of the Weyl chamber, intact. This makes it convenient to rewrite Eq.(\ref{MMIW})  in the shortcut notation used previously in Eqs. (\ref{M22}), (\ref{Malpham22})  for  $M_\alpha$ and $M_{\alpha,\,m}$, which takes into account only the components of the field $\Phi$ specified in Eq.(\ref{PhiTan}). In this notation Eq.(\ref{MMIW}) reads
\begin{equation}
M^{\,2\times 2}_{\,\alpha,\,m}\,M^{\,2\times 2}_{\,\alpha,\,m+1}
\,=\,O_{~\alpha}^{\,2\times2}\,\equiv\,
\begin{pmatrix}
-1 & ~\,\,2~ \\
~~0  &   -1~
\end{pmatrix}~.
\label{IW22}
\end{equation}
Observe that the $2\times2$ matrix $O_{~\alpha}^{\,2\times2}$ introduced here proves to be identical to the matrix $M_\infty^\text{SU(2)}$ from the Seiberg-Witten solution.
Setting $m=1$ in Eq.(\ref{MMIW}) and remembering the shortcut versions of the monodromies $M_\alpha$ and $M_{\alpha,\,1}$ from Eqs.(\ref{M=M}), (\ref{M1M-1}),  one finds that Eq.(\ref{IW22}) coincides with the identity
\begin{equation}
M_1^{\,\text{SU(2)}}\,M_{-1}^{\,\text{SU(2)}}=M_\infty^{\,\text{SU(2)}}\,,
\label{MMWSU2}
\end{equation}
which reproduces the important property of the theory with the SU(2) gauge group \cite{Seiberg:1994rs}. Eq.(\ref{MMM}) for the SU(2) gauge theory reads
$M_1^{\,\text{SU(2)}}\,M_{-1}^{\,\text{SU(2)}}\,\Omega\,M_1^{\,\text{SU(2)}}\,\Omega^{-1}=-1$.
Thus, Eqs.(\ref{MMIW}), (\ref{MMM}), which are written for the theory with an arbitrary gauge group, find a close correspondence with the theory based on the simplest SU(2) gauge group.

It is instructive to consider this set of arguments in reverse order. One can say that 
identity (\ref{MMWSU2}), which is known from the SU(2) gauge theory, confirms the validity of Eq.(\ref{MMIW}) in gauge theory with an arbitrary gauge group. The latter identity, in turn, was derived using the perturbation theory approach presented in (\ref{ADA+log+final}). Thus, one can argue that the way the perturbation theory is implemented near a wall of the Weyl chamber in Eq.(\ref{ADA+log+final}) complies with the Seiberg-Witten solution for the SU(2) gauge theory.

This circumstance is important since there exists a difference related to the perturbation theory near a wall of the Weyl chamber. This work puts forward arguments supporting  Eq.(\ref{ADA+log+final}) for the dual fields, whereas a more conventional description reads
\begin{equation}
A_\text{\it D}^{\,\prime}\,\simeq\,\frac{i}{2\pi}\,
\left(\,\hv A\, \ln\,\frac{A^2}{\Lambda^2} + (\alpha\cdot A)
\,\ln\Big(\,\frac{(\alpha\cdot A)^2}{A^2}\,\Big)~\alpha~\right)\,,
\label{ADA+log+PTh}
\end{equation}
which differs from (\ref{ADA+log+final}) by the second term; to point out this discrepancy the primed notation for the dual field $A_\text{\it D}^{\,\prime}$ is employed here. Equation (\ref{ADA+log+PTh}) follows from a conventional expression for the superpotential
in the perturbation theory
\begin{equation}
	\CF(A)\simeq \frac{i}{8\pi}\,\sum_\alpha \,(\alpha \cdot A)^2 \ln \frac{(\alpha \cdot A)^2~ }{\Lambda^2}~.
	\label{FPTh}
\end{equation}
By differentiating it over $A$ and using the large-logarithm approximation one derives Eq.(\ref{APTh2}) for the dual field. (An identity 
%\begin{equation}
$\sum_{\alpha}\,\alpha \, \otimes \,\alpha\,=\,2\,\hv$
%\label{ahv}
%\end{equation}
proves useful in this derivation.) In the vicinity of a wall of the Weyl chamber the term with one particular $\alpha$ in the sum in (\ref{FPTh}) should be treated carefully, which allows one to derive (\ref{ADA+log+PTh}) from (\ref{FPTh}).

Thus, Eqs. (\ref{ADA+log+final}) and (\ref{ADA+log+PTh}) give two different descriptions of the dual field $A_\text{\it D}$ near the wall of the Weyl chamber. The mentioned argument, which relates (\ref{ADA+log+final}) with (\ref{MMWSU2}) and therefore connects it with the known solution for the SU(2) gauge group, seems to resolve this uncertainty 
unequivocally in favour of $A_\text{\it D}$ from (\ref{ADA+log+final}) proposed in the present work. 

However, this latter fact does not necessarily indicate that there is absolutely no region  where Eq.(\ref{ADA+log+PTh}) could be applicable. Such region may exist since there exists the wall of marginal stability, which was studied in $\CN=2$ and $\CN=4$ gauge theories \cite{Argyres:1995gd,Henningson:1995hj,Bilal:1996sk,Bilal:1997st,Ritz:2000xa,Ritz:2008jf}, as well as in relation to BPS black holes in $\CN=2$ and $\CN=4$ string theories \cite{marg-wall-string}. It may happen that the wall of the Weyl chamber is separated from an interior region of this chamber by the wall of marginal stability. 
Then one can speculate that the wall of marginal stability may present a boundary separating the regions where either (\ref{ADA+log+final}) or (\ref{ADA+log+PTh}) is valid, but we will not attempt to cultivate this argument further. 

Summing up, we found that Eq. (\ref{MMIW}) brings together monodromies related to weak and strong coupling, Eq.(\ref{MMM}) expresses the operation of the weight-inversion via the monodromies that describe dyons and the dual transformation, while Eq.(\ref{OMO}) presents the quantum part of the weight-inversion via the monodromy for the monopole. The found correspondence between Eq.(\ref{MMIW}) and identity (\ref{MMWSU2}) known from the SU(2) gauge theory supports the way the perturbation theory is described in (\ref{ADA+log+final}).

\section{Degeneracy along Weyl vector}

\subsection{General approach}
\label{General-approach}
For a generic field $A$ the masses defined in Eqs.(\ref{MZ}) and (\ref{Z}) 
differ for different dyons, though by tuning $A$ one can introduce degeneracy, make some masses equal. It is shown below that there exists a particularly interesting pattern of degeneracy when the masses of {\em all} monopoles are described by the only constant.
In this case the fields $A$ and $A_\text{\it D}$ show a peculiar behaviour
being both aligned along the Weyl vector of the Cartan algebra.

Take $r$ monopoles whose magnetic charges are equal to the simple coroots of the Cartan algebra
\begin{equation}
g_i\,=\,\av_i~, \quad\quad i=1,\dots r~.
\label{giavi}
\end{equation}
Consider a particular case where masses of all these monopoles are identical, $m_{\av_i}\,=\,m$, $i=1,\dots r$.
According to Eqs.(\ref{MZ}) and (\ref{Z}) this implies that absolute values of the central charges for different monopoles are identical, $|\CZ_i|=|\av_i\cdot A_\text{\it D}|=m/\sqrt{2}$. Moreover, consider even more restrictive situation by assuming that not only the masses, but the central charges for these $r$ monopoles are identical
\footnote{Generally speaking a central charge should be assigned to a given quantum state of the theory, but since it is known from \cite{Witten:1978mh} that dyons give additive contributions to central charges one can assign the central charge to a given dyon.}
\begin{equation}
\CZ_i\,=\,\av_i\cdot A_\text{\it D}\,=\,\exp(i\phi)\,m/\sqrt{2}\,\equiv \,\CZ~.
\label{Zsame}
\end{equation}
Here $\CZ$ denotes the common central charge. Let us refer to (\ref{Zsame}) as a condition of degeneracy of central charges. We will see that this condition results in an enormous simplification of the spectrum for monopoles and dyons. Observe first of all that when this condition is satisfied, then the central charge of a monopole with an arbitrary magnetic charge $g$ can be written in a very simple form
\begin{equation}
\CZ_g\,=\,\CZ\,\sum_{i=1}^r n_i~,
\label{Zg}
\end{equation}
where integers $n_i\in Z$ define the magnetic charge
\begin{equation}
g\,=\,\sum_{i=1}^r n_i\,\av_i~.
\label{gni}
\end{equation}
Equation (\ref{Zg}) implies that the mass of a monopole with the magnetic charge $g$ has a transparent form
\begin{equation}
m_g\,=\,m\,\big| \,\sum_{i=1}^r n_i \,\big|~.
\label{m=sum}
\end{equation}
We see that the condition of degeneracy of central charges of monopoles (\ref{Zsame}) makes it certain that central charges for all monopoles are governed by only two parameters, $m$ and $\phi$, where $m$ is the only parameter that defines the masses of the monopoles and $\phi$ is the common phase for central charges related to these monopoles. 
Clearly, this degeneracy presents a vast simplification compared to a generic situation. 
Note that the derivation of (\ref{m=sum}) is based on the degeneracy of central charges formulated in (\ref{Zsame}). The mere degeneracy of masses of $r$ monopoles, i. e. identity $m_{\av_i}\,=\,m$, would not be sufficient to validate (\ref{m=sum}).

An important example of degeneracy of central charges of monopoles is the case when $m=0$ and therefore {\em all} monopoles are massless. Under this condition the ${\cal N}=2$ supersymmetric  gauge theory can be explicitly broken down to ${\cal N}=1$ theory with creation of the monopole condensate, as was discovered in \cite{Seiberg:1994rs} using an example of the SU(2) gauge theory.

Consider (\ref{Zsame}) as a set of $r$ equations on $A_\text{\it D}$. Their solution reads 
\begin{equation}
A_\text{\it D}\,=\,k_\text{\it D}\,\rho~,
\label{ADro}
\end{equation}
where $\rho$ is the Weyl vector of the Cartan algebra, and $k_\text{\it D}=
\exp(i\phi)\,|\,k_\text{\it D}\,|$ is a constant. Its absolute value defines the mass $m$ of $r$ monopoles considered, $m^2={2} \,|k_\text{\it D}\,|^2\,\rho^2$, while the phase of $k_\text{\it D}$ equals the phase $\phi$ which is present in $\CZ$ in  (\ref{Zsame}) . To verify that (\ref{ADro}) presents a solution of (\ref{Zsame}) it suffices to recall that the Weyl vector equals a sum of the fundamental weights 
\begin{equation}
\rho\,=\,\sum_{i=1}^r \,\omega_i~,
\label{ro}
\end{equation}
and hence satisfies identities
%\begin{equation}
$\av_i\cdot\rho=1$
%\label{roavi}
%\end{equation}
that follow from (\ref{avo}). 
We learn that the condition of degeneracy of central charges  (\ref{Zsame}) is equivalent to the statement that the dual scalar field is aligned with the Weyl vector (\ref{ADro}). 

A clear form of the latter result inspires one to inquire what happens with the scalar field itself when its dual satisfies (\ref{ADro}). The answer is that whenever the dual field is aligned with the Weyl vector, the scalar field is aligned with this vector as well
\begin{equation}
A\,=\,k\,\rho~,
\label{Aro}
\end{equation}
where $k$ is a complex constant. In order to justify this claim observe firstly that for weak coupling it certainly holds, as is evident from (\ref{APTh2}). To verify its validity in general case one can rely on the chiral symmetry. According to Eqs.(\ref{ma}), (\ref{monopole}) a chiral transformation converts monopoles into dyons. Since this transformation is a symmetry of the system, it necessarily preserves the pattern of degeneracy.  This implies that the spectrum of dyons, which arise from monopoles satisfies 
\begin{equation}
\CZ_i^{\,\prime} \,=\,\CZ^{\,\prime}, 
\label{Zprime}
\end{equation}
which represents a chiral transformation of Eq.(\ref{Zsame}).
Here $\CZ_i^{\,\prime}$ is a central charge of a dyon that arises under the chiral transformation from the monopole with the magnetic charge $g_i=\av_i$. 
According to  Eq.(\ref{ma}) the electric and magnetic charges of this dyon are $(g_i,q_i)=(\av_i,-m\av_i)$. Similarly, the chiral transformation converts Eq.(\ref{Zg}) into
\begin{equation}
\CZ_g^{\,\prime}\,=\,\CZ_{\,\CG}\,=\,\CZ^{\,\prime}\,\sum_{i=1}^r n_i~,
\label{Zgprime}
\end{equation}
where $\CZ_{\,\CG}$ is a central charge for a dyon whose magnetic and electric charges
arise under the chiral transformation from the monopole with the magnetic charge $g$ defined in (\ref{gni}), which means that $\CG$ for this dyon equals $\CG=(g,-mg)$. 

Using Eq.(\ref{Z}), which states $\CZ =g \cdot \AD+q\cdot A$, one finds 
from (\ref{Zprime}) that 
$\CZ_i^{\,\prime}=\CZ_i-\av_i\cdot A$. 
Equations (\ref{Zsame}) and (\ref{Zprime}) show that $\CZ_i$ and $\CZ_i^{\,\prime}$ are both $i$-independent. Consequently, $\av_i\cdot A$ is $i$-independent as well. From the latter condition we immediately derive that $A$ is aligned along the Weyl vector, precisely what Eq.(\ref{Aro}) is stating.

We learned that the degeneracy of central charges of monopoles introduced in (\ref{Zsame}) results in a number of interesting consequences. To begin with, the central charge of any monopole  satisfies very simple condition (\ref{Zg}). Moreover, for those dyons that can be created from monopoles by a chiral transformation,  
Eqs. (\ref{Zsame}) and (\ref{Zg}) are mimicked by Eqs.(\ref{Zprime}), (\ref{Zgprime}). 
This physical picture has a very clear and attractive geometrical interpretation.
The degeneracy of central charges of monopoles ensures that
the scalar field and its dual are simultaneously aligned along the Weyl vector, as Eqs. (\ref{ADro}), (\ref{Aro}) state. We can reverse the arguments, observing that from the derivation it is clear that whenever any {\it one} of the six mentioned conditions  ( Eqs. (\ref{Zsame}),(\ref{Zg}), (\ref{ADro}), (\ref{Aro}), (\ref{Zprime}), or (\ref{Zgprime})) holds, then the other five are automatically satisfied. In particular, when one of the two fields, $A$  or $A_\text{\it D}$, is aligned along the Weyl vector, then the other is also aligned along the same vector. This means that along this direction a one-dimensional description of the theory can be formulated. One can express  $k_\text{\it D}$ from (\ref{ADro}) as a function of $k$ from (\ref{Aro}). Then the variation of $k$ results in the related variation of the fields $A$ and $A_\text{\it D}$, which both remain aligned along the Weyl vector. The presented derivation of these features of the theory was based on its general properties, did not rely on the perturbation theory, being therefore valid for arbitrary coupling.

\subsection{The model}
The one-dimensional nature of the theory along the Weyl vector $\rho$, see Eqs.(\ref{ADro}) and (\ref{Aro}), can be used as a good, demanding test for the proposed model, which probes its validity at arbitrary coupling. Equation (\ref{ro}) shows that in order to align the fields $A$ and $A_\text{\it D}$ along the Weyl vector it is necessary to make $i$-independent their expansion coefficients
$A_i$ and $A_\text{\it D,\,i}$ in Eqs.(\ref{Aq}), (\ref{ADq}). From Fig. \ref{g=r} we see that this goal would be achieved provided  $z_{i,\pm}$ are $i$-independent. Let us show that this goal can be achieved by tuning the parameters $a_i$ appropriately. Assume that the coefficients $a_i$ from Eq.(\ref{ome}) satisfy
\begin{equation}
a_i\,=\,\varkappa_i\,\lambda~,
\label{akc}
\end{equation}
where $\lambda$ is a complex constant, while $\varkappa_i$ are defined by the following phase factors
\begin{equation}
\varkappa_i\,=\,\exp\left({i \,\pi  m_i\, }/{r}\right)~.
\label{km}
\end{equation}
Here integers $m_1,m_2,\dots m_r$ represent an arbitrary permutation of the sequence  $0,1,\dots r-1$ (for example $m_i=i-1$).
It is easy to verify that thus introduced factors satisfy identities
\begin{equation}
\sum_{i=1}^{r}\,(\,\varkappa_i\,)^{\,2p}\,=\,r \,\delta_{\,p,\,r}~,
%\left\{
%\begin{array}{l}
%0\,,\quad k=1,\dots r-1~,
%\\
%r\,,\quad k=r~,
%\end{array}
%\right.
\label{kp}
\end{equation}
where $p$ is an integer. Consequently, for the chosen values of $a_i$ the polynomial $P(z)$ from (\ref{YY}) is reduced to 
\begin{equation}
P(z)=z^r+(-1)^r\,\prod_{j=1}^r a_j^{2}=z^r-\lambda^{2r}~.
\label{Pz-}
\end{equation}
Deriving this equality it was noticed that Eqs. (\ref{akc}),(\ref{km}) imply that 
$\prod_{j=1}^r a_j^{2}=\lambda^{2r} \exp\big(\,(2\pi i/r)\sum_{j=0}^{r-1}j\,\big)=(-1)^{r-1}\lambda^{2r}$.
Eq.(\ref{Pz-}) allows one to simplify also  the polynomials $Y_\pm(z)$ from Eq.(\ref{YYY}) 
\begin{equation}
Y_\pm(z)\,=\,z^r -\lambda^{2r}\big( \,1 \mp ({\Lambda}/{\lambda})^{\hv}\,\big)~.
\label{Ypm1}
\end{equation}
It is taken into account 
here that Eq.(\ref{f2}),(\ref{akc}),(\ref{km}) are satisfied if one assumes that $[\,a \,]=\lambda$ (other solutions of (\ref{f2}), $[\,a \,]=\exp(\pi i n/r)\lambda$, bring do not bring new features into the problem).
%\begin{equation}
%[\,a \,]\,=\,\lambda~.
%\label{lar}
%\end{equation}
Using (\ref{Pz-}), (\ref{Ypm1}) one finds the fields from (\ref{Ai2}), (\ref{ADi2}) 
\begin{align}
&A_i~~~=\,\frac{r}{\pi }\,\int_{z_{i,\,+} }^{z_{i,\,-} }
\Big[\,\frac{  z^{2r-1} }
{ ( z_{i,\,-}^r\!-\,z^r )
( z^r - z_{i,\,+}^r) } \,\Big]^{1/2}\,dz~,
\label{AjlL}
\\
&A_{\text{\it D},\,i}\,=\,\frac{r}{\pi }\,\int_{0}^{z_{i,\,+} }
\Big[\,\frac{  z^{2r-1} }
{ ( z_{i,\,-}^r\!-\,z^r )
( z^r - z_{i,\,+}^r) } \,\Big]^{1/2}\,dz~.
\label{ADjlL}
\end{align}
Here the parameters $z_{i,\,\pm}$ satisfy conditions
\begin{equation}
(z_{i,\,\pm})^r\,=\,\lambda^{2r}\,( \,1 \mp ({\Lambda}/{\lambda})^{\hv}\,)~.
\label{zjpm}
\end{equation}
On the considered Riemann surface we can resolve them as follows
\begin{equation}
z_{i,\,\pm}\,=\,
\lambda^{2}\,( \,1 \mp ({\Lambda}/{\lambda})^{\,\hv}\,)^{\,1/r}~.
\label{zjpmlambda}
\end{equation}
Thus, we see that $z_{i,\,\pm}$ can be made $i$-independent, in accord with our expectation. 

Equation (\ref{zjpmlambda}) implies that the expansion coefficients $A_i$ and $A_{\text{\it D},\,i}$ for the scalar and dual fields in Eqs.(\ref{Aq}) and (\ref{ADq}) are independent on $i$. As was mentioned, the independence of the expansion coefficients $A_i$ and $A_{\text{\it D},\,i}$ on $i$ means that the fields $A$ and $A_{\text{\it D}}$ are both aligned along the Weyl vector $\rho$. Moreover, Eqs.(\ref{AjlL}) and (\ref{ADjlL}) express these fields as functions of a free parameter $\lambda$. This implies that a variation of $\lambda$  results in variations of $A$ and  $A_{\text{\it D}}$ under which both these fields remain aligned along $\rho$. 

Remember that this interesting feature was dubbed as the one-dimensional behaviour along the Weyl vector in Subsection \ref{General-approach}, where its existence was derived from basic  properties of the theory. The fact that the model reproduces this specific behaviour supports the validity of the model. Moreover, an absolute value of the parameter $\lambda$ is arbitrary, so by switching it from the region $|\lambda\,|<\Lambda$ to $|\lambda\,|>\Lambda$ one can  probe the model at strong and weak coupling finding that it shows this particular behaviour at arbitrary coupling, which is precisely what we wanted to verify.

\section{Analytic structure and gauge group}
\label{GG}
Generically, an arbitrary gauge group is governed by a large number of parameters. However, Eqs.(\ref{YY}) state that only two of them, the rank $r$ and the dual Coxeter number $\hv$ of the group, are present in the differential $d\lambda$. All other parameters of the gauge group appear in the theory only via the fundamental weights of the group, which define the expansion of the scalar fields in Eqs.(\ref{Aq}),(\ref{ADq}). The weights themselves include very detailed information about the group, but when the analytic structure of the solution is formulated then only the coefficients in these expansions are relevant.
These coefficients are identified with the periods in Eqs.(\ref{Ai2}),(\ref{ADi2}) and depend entirely on the properties of the differential, which is governed by the two mentioned parameters, $r$ and $\hv$. All other information related to the gauge group, including its Cartan matrix, does not manifest itself explicitly in the differential. 

Remembering that the $\tau$-matrix of the coupling constants is expressed via the expansion coefficients from Eqs.(\ref{Aq}),(\ref{ADq}) we conclude that the rank and the dual Coxeter number are the only parameters specified by the gauge group, which define the analytic structure of the model and its coupling constants.

Recall that the groups SU($r+1$) and Sp($2r$) have the same rank $r$  and same dual Coxeter number $\hv=r+1$, see Table \ref{t1} in Section \ref{differential}. Consequently, relations between the coefficients $A_{\text{\it D},\,i}$ and $A_{i}$ in these two theories are expressed by the same functions, and the $\tau$-matrices, which represent the coupling constants in these two theories, are equal
\begin{equation}
\tau^\text{\,SU(r+1)}\,=\,\tau^\text{\,Sp(2r)}~.
\label{tt}
\end{equation}
This equality may look surprising since the considered groups have very different nature, one is simply laced and the other is not, but the only fact that matters here is that $r$ and $\hv$ for them are the same.

The equality in (\ref{tt}) is valid provided the arguments in both $\tau$-matrices there are identical, $A_i^\text{\,SU(r+1)}=A_i^\text{\,Sp(2r)}$, remember that $A_i$ are the expansion coefficients for the scalar field $A$ in Eq.(\ref{Aq}). However, the equality of these coefficients does not mean the equality of the corresponding fields. On the contrary, since the fundamental weights of the gauge algebras in these two theories are different, the scalar fields derived from Eq. (\ref{Aq}) necessarily differ, $A^\text{\,SU(r+1)}\ne A^\text{\,Sp(2r)}$. Thus, repeating, Eq.(\ref{tt}) is written for one and the same set of expansion coefficients $A_i$ in Eq.(\ref{Aq}), which specifies different scalar fields for the two theories considered.

\section{Summary}
\label{Summary}
The proposed model describes the low-energy properties of the $\CN=2$ supersymmetric gauge theory for compact simple gauge groups. It is written in a universal form which is applicable to any gauge group, simply laced or not, classical or exceptional. The basis of fundamental weights, which is used in the model for the scalar fields in the Cartan algebra,  provides a convenient way to reproduce the known electric and magnetic charges of dyons \cite{Kuchiev:2008mv}. It  is  verified that the model correctly describes the region of weak coupling. Remembering that all the relevant functions in the theory are holomorphic, one can argue that since the model reproduces the boundary conditions for these functions (which are related to dyon charges at strong coupling and perturbation theory at weak coupling) it is reliable.

The model has a transparent structure. It is presented on the Riemann surface, which  has the lowest possible genus that equals the rank of the gauge group. The functional form of the differential, which is defined on this surface, is governed by only two parameters related to the gauge group, its rank $r$ and dual Coxeter number $\hv$. One could have anticipated that other parameters, which describe the gauge group, for example its Cartan matrix, may appear in the differential, but they do not, which greatly simplifies the problem.

A number of supplementary verifications to validate the model are fulfilled. It is verified that the Seiberg-Witten solution for the SU(2) gauge theory is reproduced. The weak coupling region is considered for two situations. One of them arises when the scalar field is well inside the Weyl chamber. The validity of the model in this case is verified straightforwardly. A more sophisticated situation takes place in the vicinity of a wall of the Weyl chamber. Here the monodromy is found, which relates the strong and weak coupling regimes and can be presented in a form identical to the monodromy, which exists in the Seiberg-Witten solution for the the SU(2) gauge symmetry. The match between these two monodromies supports validity of the model. The model is also shown to incorporate the important discrete transformations, namely the  chiral symmetry and duality condition. 

An additional verification of the model is based on the new interesting property of the theory. It is shown that when the vector $A$, which represents the scalar field in the Cartan algebra, is aligned along the Weyl vector $\rho$, then the vector of the dual field $A_{\text{\it D}}$ is necessarily aligned along the same direction (the opposite statement is also valid, when  $A_{\text{\it D}}$ is aligned along $\rho$, then $A$ is aligned along the same direction). 
%Thus, when both fields $A$ and $A_{\text{\it D}}$ are aligned along $\rho$, the one-dimensional description of the theory is valid. 
It is also shown that the alignment of $A$ and  $A_{\text{\it D}}$ along $\rho$ is accompanied by a profound simplification of the monopole and dyon spectra. In particular, the masses of {\it all} monopoles in this case are described by only {\it one} constant. These aspects of the theory are derived from its basic properties and remain valid for any coupling, weak or strong. The fulfilled verification shows that the model reproduces these interesting features.

The mentioned  verifications support the validity of the model. Nevertheless, since cross verifications never hurt it would be desirable to compare the present model with previous solutions. However, there is a sizable obstacle, which needs to be overtaken. In the present work the scalar fields in the Cartan algebra are presented in the basis of fundamental weights. As mentioned, this approach provides a convenient account of the known electric and magnetic charges of light dyons. Solutions considered previously have all been written using different starting points, taking the basis of simple roots or the orthonormal basis, which makes description of the dyon charges more complex, see discussion and comparison in \cite{Kuchiev:2008mv}. Thus, before matching the present model with previous solutions this complication needs to be resolved. 

The model is applied to show that the $\tau$-matrix of coupling constants for the theory based on SU($r+1$) gauge group is identical to the $\tau$-matrix for the theory with Sp($2r$) gauge group. This unexpected property of the theory stems from the fact that the pairs of parameters $r$, $\hv$ in  SU($r+1$) and Sp($2r$) groups are identical. Consequently, the relevant differentials for these groups prove to be identical as well. 
Hopefully, a clear and general nature of the model would make it useful for other applications.

%\vspace{1cm}
%\acknoledgment
A financial support of the Australian Research Council is acknowledged.

\end{document}